\documentclass[sigconf,nonacm,balance]{acmart}

\AtBeginDocument{%
  \providecommand\BibTeX{{%
    \normalfont B\kern-0.5em{\scshape i\kern-0.25em b}\kern-0.8em\TeX}}}

\usepackage{tabularx} 
\usepackage{booktabs} 
\usepackage[ruled]{algorithm2e} 
\usepackage[colorinlistoftodos]{todonotes}
\usepackage{verbatim}
\usepackage{graphicx}
\usepackage{xspace}
\usepackage{titlesec}
\usepackage[frozencache,cachedir=.]{minted}
\usepackage{tikz}

\usepackage{balance}       
\usepackage[T1]{fontenc}   
\usepackage{hyperref}

\usepackage{color}
\usepackage{booktabs}

\usepackage{microtype}        
\usepackage{ccicons}          

\usepackage{cuted}
\usepackage{capt-of}

\usepackage{upquote}
\usepackage{listings}
\usepackage{xcolor}
\usepackage{float}

\usepackage[most]{tcolorbox}

\definecolor{codegreen}{rgb}{0,0.6,0}
\definecolor{codegray}{rgb}{0.5,0.5,0.5}
\definecolor{codepurple}{rgb}{0.58,0,0.82}

\lstdefinestyle{mystyle}{
    backgroundcolor=\color{white},
    commentstyle=\color{codegreen},
    keywordstyle=\color{magenta},
    numberstyle=\tiny\color{codegray},
    stringstyle=\color{codepurple},
    basicstyle=\ttfamily\footnotesize,
    breakatwhitespace=false,
    breaklines=true,
    captionpos=b,
    keepspaces=true,
    numbers=left,
    numbersep=5pt,
    showspaces=false,
    showstringspaces=false,
    showtabs=false,
    tabsize=2
}

\definecolor{light-gray}{gray}{0.95}

\definecolor{lightgray}{rgb}{0.83, 0.83, 0.83}

\usepackage{multirow}

\usepackage{blindtext}
\usepackage{subcaption}
\usepackage{caption}

\usepackage{enumitem}
\usepackage{graphicx}
\usepackage{multicol}

\usepackage{appendix}

\usepackage{tikz}
\usetikzlibrary{positioning,arrows.meta,graphs,backgrounds,fit,calc,quotes,shapes.multipart}
\usepackage{pgfplots}
\usepackage{pgfplotstable}

\pgfplotsset{width=3.5cm,
    tick label style={font=\footnotesize}
}

\definecolor{tisanecodetop}{RGB}{74,63,106}
\definecolor{closecolor}{RGB}{225,93,87}
\definecolor{minimizecolor}{RGB}{245,215,90}
\definecolor{maximizecolor}{RGB}{111,208,74}

\SetAlFnt{\small}
\SetAlCapFnt{\small}
\SetAlCapNameFnt{\small}
\SetAlCapHSkip{0pt}
\IncMargin{-\parindent}

\newcommand{\flowco}[0]{\textsc{Flowco}\xspace}
\renewcommand{\t}[1]{\texttt{#1}}
\newcommand{\dataflow}{dataflow\xspace}
\newcommand{\dataflowCaps}{Dataflow\xspace}

\def\sketch{Flowco\xspace}

\renewcommand{\subsubsectionautorefname}{Subsubsection}

\usepackage{setspace}

\newcolumntype{H}{>{\setbox0=\hbox\bgroup}c<{\egroup}@{}}

\makeatletter
\newcommand\verytiny{\@setfontsize\verytiny{5pt}{6pt}} 
\makeatother

\def\<#1>{\codeid{#1}}
\newcommand{\codeid}[1]{\ifmmode{\mbox{\small\ttfamily{#1}}}\else{\small\ttfamily #1}\fi}
\newcommand{\codeidsmall}[1]{\ifmmode{\mbox{\smaller\ttfamily{#1}}}\else{\smaller\ttfamily #1}\fi}

\NewTColorBox{mybox}{ O{red} m d"" !O{} }{
    enhanced,
    colframe=#1!25!white,
    colback=#1!5!white,
    coltitle=black,
    fonttitle=\bfseries,
    title=#2,
    beforeafter skip=0.25em,
    top=0em,
    bottom=0em,
    left=0.25em,
    right=0.25em,
    boxrule=0.5pt
}

\NewTColorBox{amabox}{ O{red} m d"" !O{} }{
    enhanced,
    colframe=gray!25!white,
    colback=gray!5!white,
    coltitle=black,
    fonttitle=\bfseries,
    title=#1,
    beforeafter skip=0.25em,
    top=0em,
    bottom=0em,
    left=0.25em,
    right=0.25em,
    boxrule=0.5pt
}

\newcommand{\analyst}[0]{Alex\xspace}

\setcopyright{rightsretained}
\copyrightyear{2025}

\begin{document}

\title{\sketch: Rethinking Data Analysis in the Age of LLMs}

\author{Stephen N. Freund}
\email{freund@cs.williams.edu}
\orcid{0009-0000-6992-199X}
\affiliation{%
  \institution{Williams College}
  \country{Williamstown, MA, USA}
}

\author{Brooke Simon}
\email{bsimon2000@g.ucla.edu}
\orcid{}
\affiliation{%
  \institution{University of California, Los Angeles}
  \country{Los Angeles, CA, USA}
}

\author{Emery D. Berger}
\authornote{Work done at the University of Massachusetts Amherst.}
\email{emery@cs.umass.edu}
\orcid{}
\affiliation{%
  \institution{University of Massachusetts Amherst / Amazon Web Services}
  \country{Amherst, MA, USA}
}

\author{Eunice Jun}
\email{emjun@cs.ucla.edu}
\orcid{0000-0002-4050-4284}
\affiliation{%
  \institution{University of California, Los Angeles}
  \country{Los Angeles, CA, USA}
}

\renewcommand{\shortauthors}{}

\begin{abstract}

Conducting data analysis typically involves authoring code to transform,
visualize, analyze, and interpret data. Large language models (LLMs) are now
capable of generating such code for simple, routine analyses. LLMs promise to
democratize data science by enabling those with limited programming expertise to
conduct data analyses, including in scientific research, business, and
policymaking. However, analysts in many real-world settings must often exercise
fine-grained control over specific analysis steps, verify intermediate results
explicitly, and iteratively refine their analytical approaches.  Such tasks
present barriers to building robust and reproducible analyses using LLMs alone
or even in conjunction with existing authoring tools (e.g., computational notebooks).  
This paper introduces \flowco, a new mixed-initiative
system to address these challenges. \flowco leverages a visual dataflow programming
model and integrates LLMs into every phase of the authoring process.  A user
study suggests that \flowco supports analysts, particularly those with less
programming experience, in quickly authoring, debugging, and refining data
analyses.

\end{abstract}

\begin{teaserfigure}
  \centering
  \includegraphics[width=\textwidth]{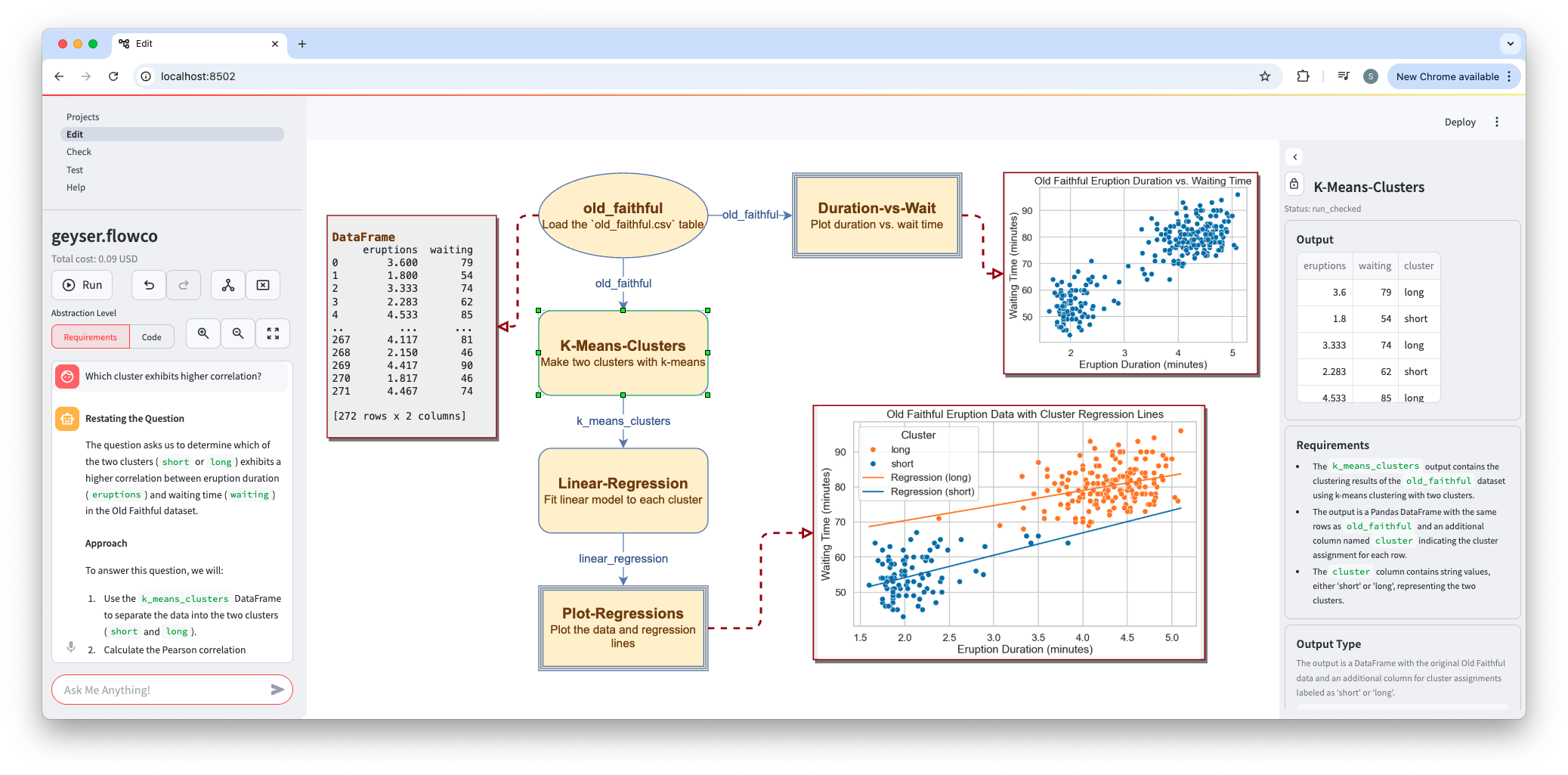}
  \caption{\flowco is a mixed-initiative system that leverages a visual dataflow programming model and 
  LLMs to not just synthesize code but also assist in all stages of development.  In this example, 
  the analyst creates a workflow to read Old Faithful geyser data, clusters eruptions by duration, and
  fits a linear model to each cluster.  \flowco translates this graph into executable code via an LLM and exposes
  a variety of other interactions to assist the analyst, including (i) a chat box with direct access to the
  data, graph, and outputs as well as (ii) user-defined checks to validate generated code.}
  \Description{
    A screenshot of the Flowco interface showing a visual data analysis dataflow graph for the Old Faithful geyser dataset. The canvas shows a dataflow graph including nodes for loading the dataset, performing K-means clustering into two groups, fitting linear regression models, and plotting results. A dataframe with columns `eruptions' and `waiting' and two scatter plots are also displayed in the canvas.  One plot shows eruption duration vs. waiting time, and the other shows the same data color-coded by cluster, with regression lines overlaid. The sidebar includes a question prompt about which cluster has a higher correlation, along with a step-by-step explanation produced by an LLM.  The explanation also includes the output table showing cluster assignments.
    }
  \label{fig:geyser}
\end{teaserfigure}

\maketitle

\renewcommand{\subsectionautorefname}{section}
\renewcommand{\subsubsectionautorefname}{section}
\renewcommand{\paragraphautorefname}{section}

\section{Introduction}

\begin{figure}
    \includegraphics[width=0.40\textwidth]{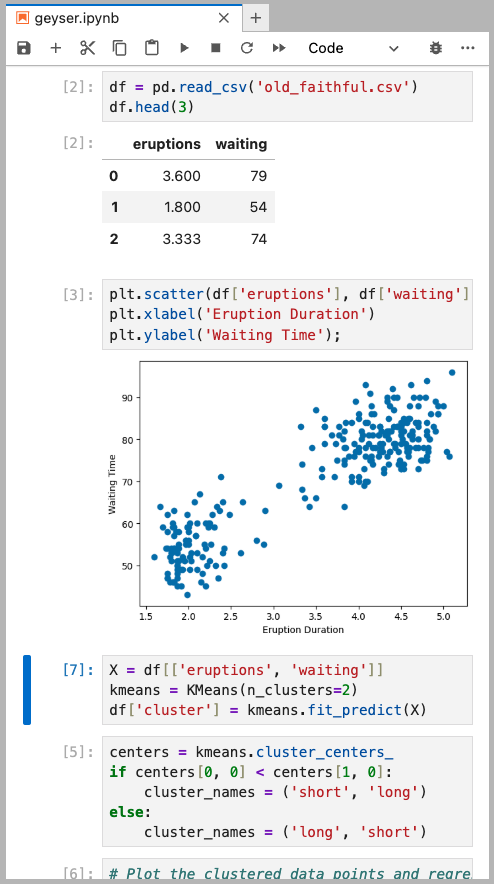}
    \caption{\label{fig:nb} A Jupyter notebook exhibiting a
      potentially stale variable.  To the left of each cell is the execution count, 
      which indicates the order in which the cells were evaluated.
      The cell labeled with execution count 7 (highlighted)
      was evaluated after \texttt{cluster\_names} was initialized in
      the cell with execution count of 5.  The latter depends on the former, meaning
      that \texttt{cluster\_names} could be stale.}
    \Description{
      A screenshot of a Jupyter Notebook showing code and visual output for analyzing the Old Faithful geyser dataset. The notebook imports data from old_faithful.csv and displays the first three rows with columns eruptions and waiting. A scatter plot shows eruption duration on the x-axis and waiting time on the y-axis, revealing two visible clusters. Below the plot, a code cell uses the K-Means algorithm with two clusters to assign cluster labels to the data. A conditional block assigns human-readable names ``short'' and ``long'' to the clusters based on their center positions.
    }
\end{figure}

Data analysis often involves incrementally transforming and
visualizing data, posing hypotheses, and applying statistical methods
through highly iterative exploratory workflows.
Computational notebooks support these workflows, but with some
limitations and challenges~\cite{kery2018storyNotebook,rule2018exploration}, and large
language models (LLMs) have the potential to drive the creation of
even more robust and lower-overhead analysis authoring tools.  However, they
too have limitations that must be carefully addressed to ensure successful
outcomes in data analysis authoring workflows.  This paper presents a
new approach for supporting mixed-initiative authoring of data
analyses with LLMs that overcomes those obstacles.

\subsubsection*{Computational notebooks.}
Figure~\ref{fig:nb} shows a Jupyter notebook that loads geyser
eruption data~\cite{geyser}, plots the raw data, and then clusters
into two groups in preparation for a linear regression analysis.

The very same features that make notebooks effective for exploratory
programming also lead to a collection of well-known impediments to
building robust workflows in
them~\cite{kery2018storyNotebook,rule2018exploration}.  They lack
modularity and abstraction, making it challenging to reason about the
overall structure of a notebook, track dependencies across cells, and
reproduce results.  These issues become particularly salient as the
number of cells grows and their contents evolve. Further, notebooks
are ill-suited for the standard unit testing and validation practices
used to help provide assurances about the quality of code in
traditional software engineering~\cite{quaranta2022collabNotebooks}.

The potential for non-linear execution order and stale values can lead
to confusion and errors in analysis that are difficult to diagnose and
fix.  For example, in Figure~\ref{fig:nb}, the cell labeled with execution count 7
(highlighted) was evaluated after \texttt{cluster\_names} was
initialized in the cell with execution count 5.  The latter depends on the
former, meaning that \texttt{cluster\_names} could be stale.

Many tools and services aim to recover
dependencies~\cite{DBLP:conf/chi/HeadHBDD19,DBLP:journals/pvldb/ShankarMCHP22}
and address other shortcomings of
notebooks~\cite{rawn2025pagebreaks,wenskovitch2019albireo}.  These
tools help mitigate many challenges of using notebooks, but they do
not fully address the need for modularity and abstraction in data
science. Modularity and abstraction are necessary to ensure the
reproducibility of data analyses throughout development.

\subsubsection*{Large language models.}
At the same time, large language models (LLMs) present opportunities
to quickly author, debug, and iterate on code. LLMs also present new
challenges: authoring effective prompts, detecting issues in
generated code, and managing possibly drastic differences in the code
produced in response to repeated queries or requests for small
modifications, among others.

Consider a scientist attempting to use an LLM to generate the above
data analysis workflow.  They start by asking the LLM to write a
script to read in the geyser data.  After creating that code, the scientist
inspects the first few rows of the data and and then asks the LLM to
plot the data as a scatter plot.  Seeing the distribution of points,
the scientist proceeds to prompt the LLM to write code to cluster the
data and compute a linear regression model for each cluster.  Running
the resulting code results in the error
\begin{verbatim}
X does not have valid feature names, but LinearRegression 
was fitted with feature names.
\end{verbatim}
because the regression model was fitted on a pandas DataFrame (which
has column names), but predictions are being made on a plain NumPy
array in the LLM-generated code.  The scientist asks the LLM to fix the
error, and also asks the LLM to use the names ``short'' and ``long'' to
distinguish the two clusters.

After making those changes, the scientist runs the code and sees that
the plot is not as expected.  The LLM has plotted waiting time
on the x-axis and eruption duration on the y-axis, which is the
opposite of what the analyst wants.  Inspecting the code, they
discover the regression was performed on the wrong variables.  This
time, the scientist finds it easier to directly add a line of code
enforcing categories and now asks the LLM to update the rest of the
code to match that edit.  While the LLM provides an executable script
after a few attempts, the scientist remains unsure of the quality of
the code or validity of the results.

\subsubsection*{Mixed-initiative authoring with LLMs for data science.}
There are many current attempts to combine LLMs and computational
notebooks in order to amplify their
strengths~\cite{anaconda_assistant,databricks_assistant,jupyter_ai,openai_data_analysis}.
These tools focus on incorporating chat interfaces into notebooks.
While potentially beneficial in some cases, such use of LLMs can
exacerbate the challenges of producing modular, robust, easily
understood, and testable data science workflows.  Moreover, these
tools do not eliminate the need for programmers to reason about the
code they are writing in a notebook.

\subsubsection*{\flowco.}
This paper introduces \flowco, a new mixed-initiative system to
address these challenges when authoring data analyses. \flowco
leverages a visual \dataflow programming model and LLMs at every phase
of the authoring process.  Users draw \dataflow graphs in which nodes
specify concrete steps in an analysis and edges indicate the flow of
information.  The \flowco implementation of the geyser analysis is
shown in Figure~\ref{fig:geyser}.  
\flowco leverages the classic dataflow model for two key reasons:
\begin{itemize}
\item Dataflow graphs provide a natural medium for exploration,
  design, and communication of data analysis workflows
  (\autoref{sec:dataflowGraphs}).  This insight comes from our
  observations about external artifacts scientists already rely upon
  to plan, discuss, and share their data analyses.
\item Dataflow graphs enjoy strong modularity, abstraction, and
  composition properties.  Those features not only support the
  creation of robust, reusable, and reproducible data analysis
  workflows from a programming perspective but also provide a natural
  representation of the computation and its components when
  interacting with LLMs.
\end{itemize}

\flowco translates \dataflow graphs into executable code via an LLM
and exposes a variety of other interactions to give the analyst
fine-grained control.  These include ways to edit individual nodes at
different levels of abstraction; an ``Ask Me Anything!'' (AMA) chat
agent with direct access to the data, graph, and outputs; and
assertion checking and unit testing capabilities to gain confidence in
the generated code and analysis.

We perform two different evaluations.  First, we demonstrate the
applicability of \flowco{} over a diverse range of analysis goals and
complexity.  Second, we conduct a user study with 12 students familiar
with data science but with a range of programming experience.  We find
that these analysts are able to successfully use \flowco{}, find the
dataflow programming model helpful for their organization and
visualization of multi-step analyses, and preferable to using ChatGPT
to perform similar tasks. Participants also found \flowco{} most
useful for getting started with analyses, especially for those with little
programming experience.

This paper contributes the following:
\begin{itemize}
\item a \dataflow programming model featuring multiple abstraction
  layers that serves as a foundation for reliable LLM-centric
  programming;
\item \flowco, a system for authoring data analysis workflows with
  LLM assistance at every stage of the process;
\item a collection of fully automatic and user-guided LLM-based
  techniques for designing, implementing, validating the
  correctness \flowco \dataflow graphs; and
\item validation that \flowco can effectively support data
  analysis authoring tasks, as demonstrated by examples and a user
  study with twelve data science students.
\end{itemize}

\section{Background and Related Work}

\subsection{Computational Notebooks for Data Science}

Computational notebooks, most notably Jupyter Notebook~\cite{jupyter}, are
widely used programming environments for data analysis. Notebooks support rapid
iteration and exploration, where analysts generate
many code versions as they work through implementation details before converging
on a design and implementation (``expand-then-reduce'')~\cite{kery2018storyNotebook}. 

However, notebooks can very quickly become complex and difficult to maintain. For
example, Kery et al. found that while notebooks do not inherently limit the
document length (i.e., number of cells), analysts run into practical length
limitations when scrolling across cells to run them in specific orders~\cite{kery2018storyNotebook}. A common practice among notebook users
is to create new notebooks that contain only the most important cells after
initial stages of exploration~\cite{kery2018storyNotebook,
dong2021cleaningNotebooks}. Interestingly, Dong et al. found that modularizing code by
creating functions and classes was the least common ``cleaning'' activity among
data scientists using notebooks~\cite{dong2021cleaningNotebooks}. A key tension
is that notebooks are primarily helpful for getting started with and iterating
on programs rapidly, but analysts quickly face code
maintenance issues that require more involved strategies (e.g., modularization,
composition, documentation) that would take them out of exploration~\cite{rule2018exploration}.

Previous work has contributed several ideas and
systems for improving code quality without compromising exploration in notebooks: extracting ``clean'' notebooks
through program slicing~\cite{head2019managingMesses}; grouping and hiding cells in a notebook
through ``cell folding''~\cite{rule2018cellFolding}; grouping and annotating cells to support
sensemaking~\cite{DBLP:conf/chi/ChattopadhyayPH20}; laying cells out two-dimensionally~\cite{harden2023twoDNotebooks};
end-user-defined scoping of variables inside notebooks~\cite{rawn2025pagebreaks}; and
visualizing cell dependencies in a notebook~\cite{wenskovitch2019albireo}. 
\flowco is most similar to prior work on dataflow notebooks~\cite{koop2017dataflowNotebooks}. 
While Wenskovitch et al.
generate dataflow graphs from notebooks to help users understand their code, \flowco gives users direct control over the dataflow graphs. 
Unlike dataflow notebooks, \flowco
integrates programming support from LLMs. Furthermore, in dataflow notebooks,
the data scientist still encounters the challenges of maintaining dependencies
during rapid exploration, which may result in multiple versions of the same
cell, all of which are represented as different dataflow paths. 
However, with
\flowco, the structure of the dataflow graph captures all dependencies between computation steps, relieving the user from the burden of reasoning about them. Finally,
\flowco can also generate a notebook from the dataflow graph.

\subsection{Interacting with LLMs}

LLMs can further accelerate programming through code generation capabilities, but they come with their own challenges.
Recent work has observed that structured problem decomposition and generation 
can reduce the metacognitive demands and challenges of using generative
AI~\cite{DBLP:conf/chi/TankelevitchKSS24}.
One technique to decompose problems is through chaining.
Chaining has become a standard
technique for decomposing complex tasks into subtasks for LLMs to
tackle~\cite{wu2022aiChains}. Wu et al. find that decomposing and chaining tasks leads to
improved performance and completion success~\cite{wu2022aiChains}. 

Building on this work, Arawjo et al. develop
ChainForge~\cite{arawjo2024chainforge}, a system for constructing prompts and
evaluating the outputs of LLMs. ChainForge provides a visual programming
interface that is similar to a dataflow graph. Arawjo et al. report ``surprise'' that
ChainForge users wanted to use ChainForge to specify data analysis pipelines, which they could not easily do. 
This paper explores this finding further. 
\flowco also goes beyond this prior work to develop
techniques for ensuring reliability of LLM-generated code in the context of data
science. Specifically, \flowco provides users with multiple layers of
abstraction, automatic change detection and propagation, and behavioral
validation to control and understand LLM code generation. 
While ChainForge
focuses on providing user support for prompt engineering, \flowco is focused on
using LLMs for data analysis. 

CoLadder introduces a hierarchical approach to decomposing programmer
intent into sub-goals~\cite{yen2024coladder}. An LLM can address each
sub-goal. While our goals are similar in supporting intent
decomposition and code composition, there are a few major
distinctions: (1) Unlike \flowco, CoLadder does not provide any
architecture for ensuring that the composed code is executable or correct
programmatically. (2) While CoLadder guides users through task
decomposition in a domain-agnostic way, \flowco takes advantage of
dataflow graphs to describe task decomposition. Notably, Figure
12 in the CoLadder paper is similar to a dataflow graph. By focusing on
dataflow graphs, we hypothesize that our work is approachable to data scientists
across many different levels of expertise.
While not focused on programming with creating inputs to LLMs or programming
with LLMS, Graphologue~\cite{jiang2023graphologue} supports decomposition of LLM outputs.
Similarly, Sensescape~\cite{suh2023sensecape} supports visual and hierarchical
exploration of LLM outputs in order to support information foraging and sensemaking. 

The multiple layers of abstraction that \flowco provides are high-level descriptions, requirements, and code. 
Zamfirescu-Pereira et al.'s PAIL IDE~\cite{zamfirescu2025pail} explores similar layers of abstraction to support users in designing programs. 
While PAIL is focused on surfacing requirements, design decisions, and code as a way to guide code generation and summarization, 
\flowco uses these layers of abstraction in order to generate and validate code (\autoref{sec:model}).

More specifically in the data analysis space, Kazemitabaar et al. explore two
different approaches for integrating LLMs into authoring data analyses with a
focus on validation: phasewise and stepwise steering~\cite{DBLP:conf/uist/Kazemitabaar0DG24}. The authors conclude by
outlining the need for interaction flows and affordances for steering LLMs. \flowco realizes this.

While prior work has given users more fine-grained control over the process of programming with LLMs, 
several limitations persist:
(1) the absence of assurances that the generated code is correct
programmatically, meaning connected pieces of code are not guaranteed
to execute (e.g., CoLadder);
(2) the lack of compositionality of code fragments generated in
different ``runs''; and
(3) absence of incremental (re)computation when
modifications occur in parts of the LLM-generated program.

\subsection{Combining Computational Notebooks and LLMs for Data Science}

The recent proliferation of code generated by LLMs has also made data science
programming more accessible. Given the wide adoption of computational notebooks
and the ability to generate code with LLMs quickly, 
how to combine both computational notebooks and LLMs for data
science is an open research topic and the focus of many emerging
products~\cite{anaconda_assistant,databricks_assistant,jupyter_ai,openai_data_analysis}.

However, the use of LLMs can further exacerbate the challenges
of designing robust, modular, and well-test data analyses.
Indeed, scientists relying on current conversational interfaces to
LLMs must critically assess whether the generated code is correct
and aligns with their scientific objectives to avoid errors that threaten statistical 
and scientific validity, a daunting task for
even expert programmers~\cite{chopra2023conversational}. 

An alternative approach to putting LLMs into notebooks and notebooks into LLMs
is necessary. Our goal is to maintain the benefits of rapid code generation,
program exploration, and iteration while also mitigating the issues of
modularity, composability, and extensibility. In response, we develop \flowco
to prioritize these qualities. A key idea in \flowco is to design for these
qualities by leveraging a data structure common in data science
and emerging as beneficial in the interactive LLMs literature: dataflow graphs.

\subsection{Dataflow Graphs} \label{sec:dataflowGraphs}

In our previous observations of and collaborations with scientists, we found
that scientists outline key computational steps, connect steps that depend on each other, and refine the steps and connections. 
How scientists describe, formulate, and communicate their analyses is 
similar to how one describes and structures \dataflow graphs.

Dataflow graphs and programming models are a classic concept in
computing with a long and rich history dating back to the 1960s and
1970s~\cite{DBLP:journals/csur/JohnstonHM04}.
Dryad~\cite{DBLP:conf/eurosys/IsardBYBF07},  
Naiad~\cite{DBLP:conf/sosp/MurrayMIIBA13}, and 
TensorFlow~\cite{DBLP:conf/osdi/AbadiBCCDDDGIIK16}, as well as the widely used
LabVIEW~\cite{DBLP:journals/pacmpl/Kodosky20} visual programming language, 
are among the many prominent
contemporary examples of dataflow-based systems.  

Visual programming dataflow languages enable users to  
organize complex computations into manageable, independently understandable, and
composable components.  Further, dataflow graphs provide a framework well-suited for
exploratory programming, where users can quickly prototype and test ideas by
interactively modifying the graph and observing the results~\cite{DBLP:journals/csur/JohnstonHM04}.  
A variety of data science and data analytics tools embrace visual dataflow graph
programming models for these reasons.  Representative examples include
Texera~\cite{DBLP:journals/pvldb/WangHNKALLDL24}, Alteryx~\cite{alteryx}, KNIME
Analytics Platform~\cite{knime}, and Altair RapidMiner~\cite{altair}.  
There have also been efforts to extract dataflow graphs from existing code~\cite{DBLP:journals/ibmrd/PattersonMSBMV17}.

\dataflowCaps graphs in \flowco provide the programming model
for organizing, defining, and reasoning about analyses.  Further, \dataflow
graphs provide a natural framework for facilitating the creation, composition,
testing, and maintenance of code generated by LLMs.  \dataflowCaps graphs are thus
not only a useful model for the user but also serve as a foundation for
reliable LLM-centric programming.  That insight is a key contribution of
\flowco.

\newcommand{\step}[1]{%
\protect
\tikz[baseline=(X.base)] 
\node (X) [draw, shape=circle, inner sep=0, fill={rgb,255:red,85; green,188;
blue,190}, text=white,draw=none] {\strut \normalfont \sffamily #1};}

\newcommand{\node}[1]{\protect\textbf{#1}}

\begin{figure*}[t!]
  {\includegraphics[width=\textwidth]{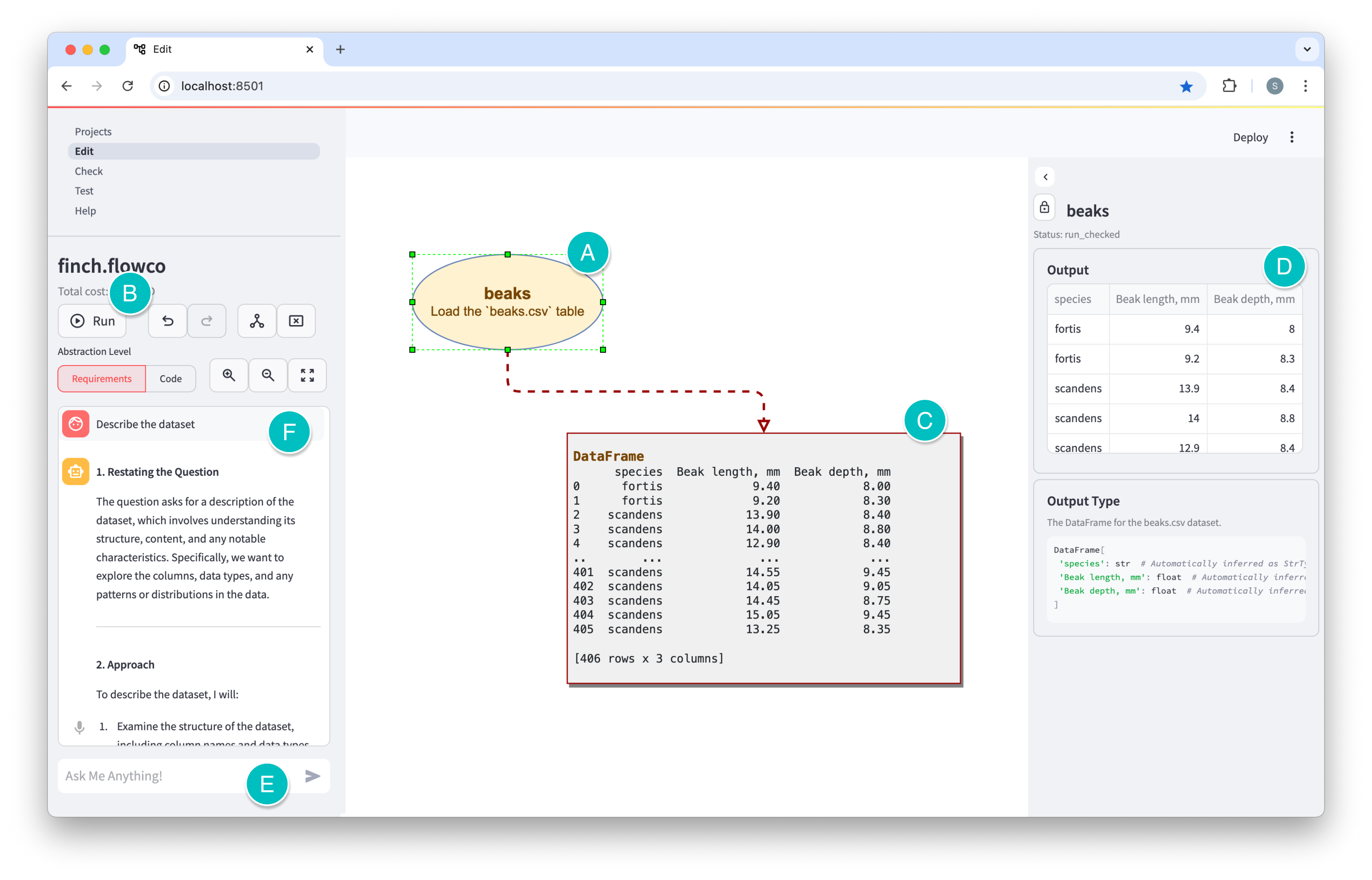}}
  \caption{\label{fig:ui}The \flowco editor interface is divided into three
  panels:  \textbf{(Left)} the \emph{project panel} encompasses global actions
  and the ``Ask Me Anything!'' (AMA) chat box;
  \textbf{(Center)} the \emph{canvas} is the visual editor for \flowco \dataflow
  graphs; and
  \textbf{(Right)} the \emph{details panel} presents details of the selected
  node during editing.
  \step{A} The user creates a new node to load
  the dataset in \texttt{beaks.csv}. 
  \step{B} The user presses the Run button
  to synthesize code to evaluate the \dataflow graph. 
  \step{C} After evaluating the node, \flowco provides a sample of the dataset
  in the canvas.  
  \step{D} The user examines
  the full dataset in the details panel.
  \step{E} The user prompts the AMA chat box to ``Describe the dataset''.
  \step{F} \flowco responds as it performs a number of analyses on the dataset.
}
\Description{
A screenshot of the Flowco interface showing a visual data analysis dataflow
graph for the initial steps of analyzing the finch dataset. The canvas contains
one node for loading the dataset and its output.  The output is a dataframe with
columns for species, beak length, and beak depth. The project panel sidebar on
the left includes several controls, including a Run button, and an ``Ask Me
Anything!'' (AMA) chat box. The user has typed a question into the AMA chat box,
asking Flowco to describe the dataset. The response from Flowco is shown.  The
details panel on the right shows the output of the selected node and the type of
output produced by that node.  
UI elements corresponding to specific steps in the usage scenario are
highlighted with colored circles labeled `A' to `F'.
}
\end{figure*}

\begin{figure*}[tp!]
  {\includegraphics[width=\textwidth]{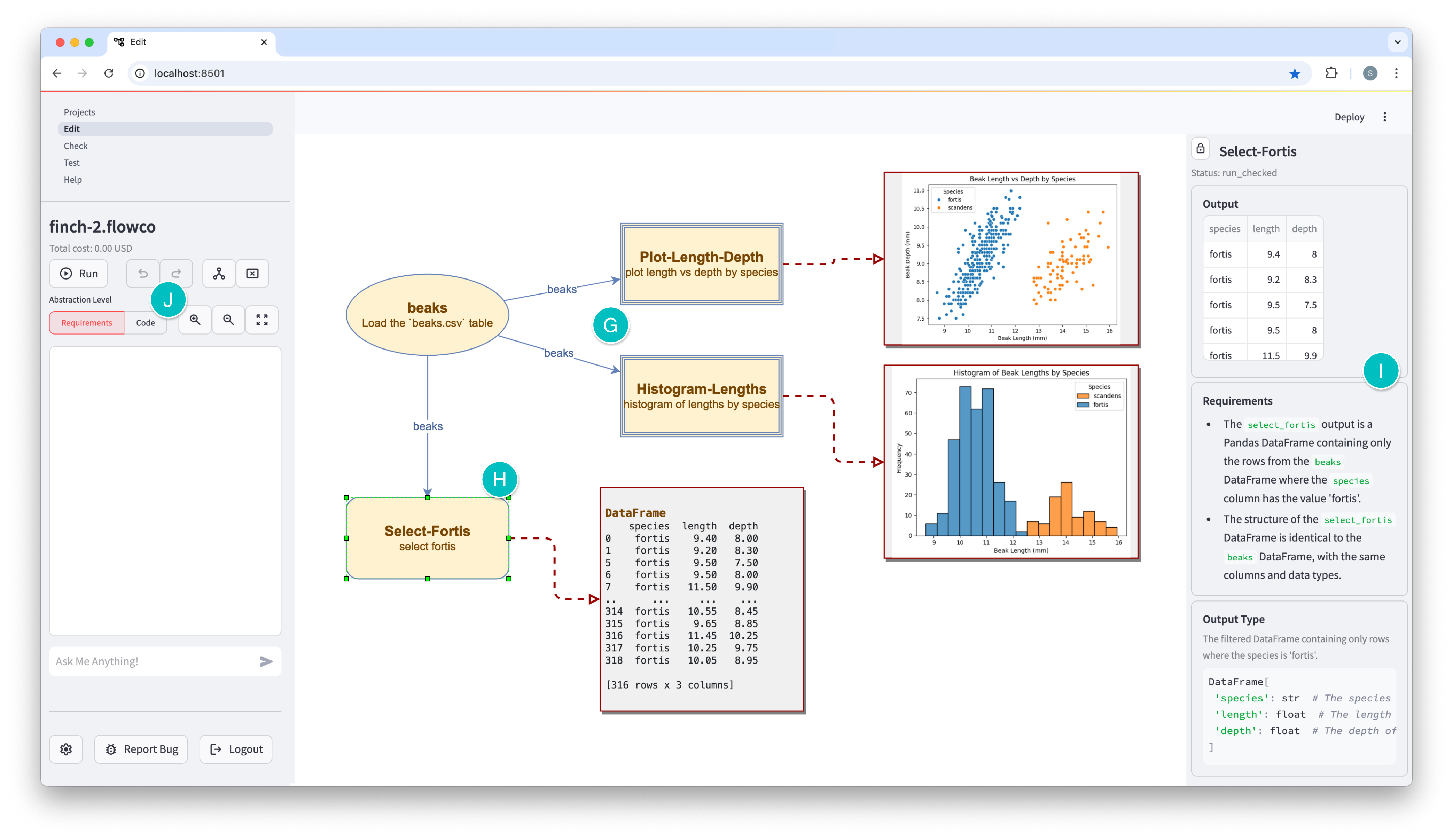}}
  \caption{\label{fig:uitwo}
  \step{G} The user adds two plotting nodes to the graph, as well 
  as \step{H} a node to select only the Fortis finches from the dataset.
  After running the graph, \step{I} the user selects \node{Select-Fortis} to examine
  it in the details panel.  \step{J} The user exposes the synthesized code
  by selecting the ``Code'' abstraction level.
  }
  \Description{
  A screenshot of the Flowco interface showing a partially-completed visual data
  analysis dataflow graph for the finch dataset. The canvas shows a node for
  loading the data set as well as three connected nodes to make a scatter plot
  of beak depth vs. length, to make a histogram of beak lengths, and to select
  only the Fortis finches from the dataset.  The details of the node to select
  Fortis finches is shown in the details panel on the right.  The user has
  selected the ``Code'' abstraction level from the project panel, which exposes
  the synthesized code for that node.
  UI elements corresponding to specific steps in the usage scenario are
  highlighted with colored circles labeled `G' to `J'. 
  }
\end{figure*}

\begin{figure*}[t!]
  \includegraphics[width=\textwidth]{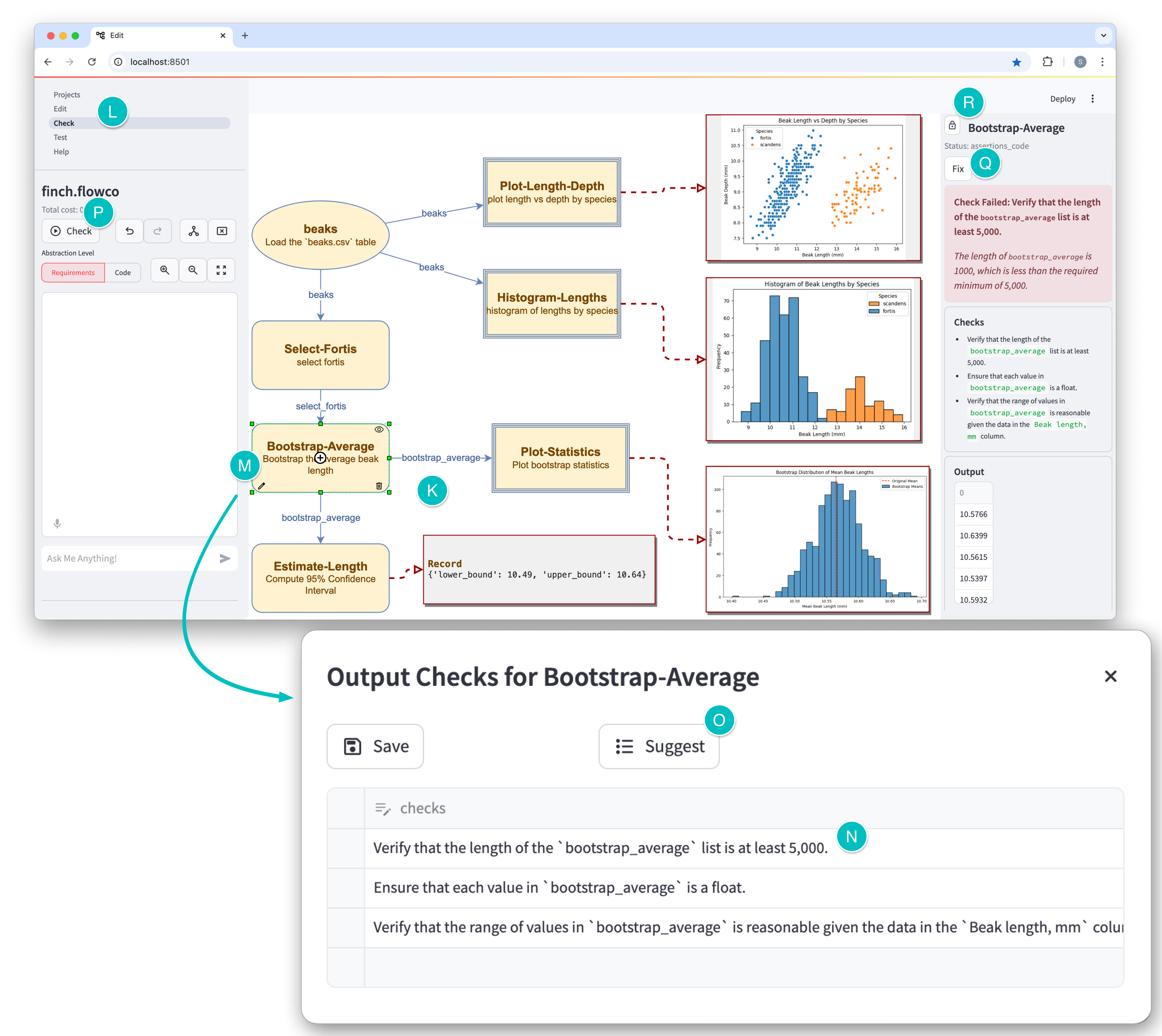}
  \caption{\label{fig:checks}\flowco enables the user to validate
    run-time assertion checks on node outputs via the Checks view.
    After \step{K} adding nodes to estimate the mean beak length for
    the Fortis finches \step{L} the user switches to the Checks view
    and \step{M} clicks the pencil icon that appears while hovering
    over \node{Bootstrap-Average} to bring up the dialog box shown below the screenshot.
   \step{N} The user manually adds the check ``Verify that the length
    of the \texttt{bootstrap\_average} list is at least 5,000", and
    then \step{O} clicks the Suggest button to have \flowco suggest
    several additional checks for that node.  After saving the checks
    the dialog closes and \step{P} the user then clicks the Check
    button to verify that all checks pass.  \step{Q} \flowco reports
    a failure for the \node{Bootstrap-Average} node. }
    \Description{
    A screenshot of the Flowco interface showing the complete visual data
    analysis dataflow graph for exploring the finch dataset. The canvas shows a
    node for loading the data set as well as three connected nodes to make a
    scatter plot of beak depth vs. length, to make a histogram of beak lengths,
    and to select only the Fortis finches from the dataset.  The selection node
    is connected to a node to bootstrap the mean Fortis finch beak length.  That
    node is then connected to nodes to show a histogram of the bootstrap
    statistics and to estimate the mean beak length as a confidence interval.
     
    The user has selected the Checks view from the project panel on the
    left.  The user has also clicked on the pencil icon to edit the checks for
    the Bootstrap-Average node.  The dialog box, shown at the bottom of the
    figure shows a list of checks for that node.
 
    The details of the node Bootstrap-Average is shown in the details panel to
    the right in the main interface.  The details include a message about a
    failing check and a button to attempt an automatic repair.

    UI elements corresponding to specific steps in the usage scenario are
    highlighted with colored circles labeled `K' to `Q'.
    }
\end{figure*}

\section{Usage Scenario} 
\label{sec:usage-scenario} 

The \flowco user interface, as shown in Figure~\ref{fig:ui}, is is
divided into three panels: The \emph{project panel} on the left
encompasses global actions and the ``Ask Me Anything!'' (AMA) chat box;
the \emph{canvas} in the center is the visual editor for \flowco
\dataflow graphs; and the \emph{details panel} on the right presents
details of the selected node during editing.  The LLM is an integral
part of \flowco, which uses it to support data exploration, \dataflow
graph creation and editing, code synthesis, validation, and error
detection and recovery\footnote{\flowco's default LLM is OpenAI's
  GPT-4o~\cite{gpt-4o}.  That model supports a number of features \flowco leverages,
  including multi-modal inputs, tool calls, streaming, and structured output.}.

The following usage scenario illustrates the most salient aspects of \flowco's
design.  A researcher, \analyst, conducts exploratory,
confirmatory, and estimation analyses on a dataset containing beak length and
beak depth measurements for two species of finches, Fortis and
Scandens~\footnote{ The data is drawn from a larger
dataset~\cite{finchdata} gathered during a long-running study of finches on Daphne
Major Island, Ecuador~\cite{grant2014forty}. }.
\paragraph*{Exploring the Data.}
\analyst begins by clicking on the initial blank canvas to create a new node.
\step{A} \analyst specifies via the node creation dialog that the node should load
a dataset from the file \texttt{beaks.csv}. \step{B} \analyst then clicks the
Run button. \step{C} \flowco then synthesizes requirements and Python code for
that node and then runs it, after which \flowco provides a sample of the dataset
in the canvas.  \analyst selects the node and \step{D} examines
the full dataset in the navigable view presented to the right in the details panel.
That panel also includes the type of output produced by that node.  In this
case, the output is a \texttt{DataFrame} with columns for species, beak length, and beak
depth.

After inspecting the output, \step{E} \analyst uses the
AMA chat box and asks that \flowco ``Describe the dataset.'' \flowco leverages its
underlying LLM to answer that question.  The LLM utilizes \flowco to run code it 
creates to examine the shape and column types for the
dataset, to verify that no values are missing, and to compute basic descriptive
statistics for each numerical column.  \step{F} The LLM reports on those steps
in its response. The LLM even runs code to generate histograms for the numerical
columns and, leveraging its multi-modal capabilities, visually inspects those
plots to conclude for beak lengths:
\begin{quote}
  \emph{The distribution appears bimodal, with two peaks around 11 mm and 14 mm. This
  suggests the presence of two distinct groups or species with different beak
  lengths.}
\end{quote}
and for beak depths:
\begin{quote}
  \emph{The distribution is unimodal and roughly symmetric, centered around 9 mm.
  }
\end{quote}

\paragraph*{Visualizations.}
Equipped with this knowledge, \step{G} \analyst adds two more nodes, each of which
will produce a plot.\footnote{In \flowco, the shape of a node indicates its
behavior: an oval represents a data-loading node, double-bordered rectangle
represents a plotting node, and a rounded rectangle represents a computation
node.} The node \node{Histogram-Lengths} generates a histogram of the beak
lengths, and the node \node{Plot-Length-Depth} plots beak length vs. depth.  The
`Run' command again synthesizes requirements (using the structure of the graph,
the requirements of predecessor nodes, and the labels to infer intent), Python
code, and finally the computed output. 
After seeing the initial version of those plots, \analyst instructs \flowco
to ``Use different colors for different species'' via the AMA chat box.  The LLM
updates the plotting nodes to reflect that request, leading to the final
versions shown in Figure~\ref{fig:uitwo}.

\paragraph*{Computation and Analysis.}
After inspecting the plots, \analyst asks via chat whether the difference in beak
lengths between the two species is statistically significant.  As before, the LLM
leverages its ability to run code and employs a Kruskal-Wallis hypothesis test
to conclude that the difference is statistically significant with a p-value of
$1.5 \times 10^{-47}$.

\analyst decides to next focus on the Fortis finch species and estimate the
average beak length for that species using the sample in the dataset.  To do so,
\step{H} \analyst adds the computation node \node{Select-Fortis} to select only
the rows for Fortis finches from the dataset, clicks Run, and \step{I} then
selects the new node to examine the node in the details panel. For computation
nodes, the details panel shows not just the output and output type, but also the
requirements for the node's behavior inferred by \flowco.

They decide to inspect the code for it
by \step{J} selecting the ``Code" abstraction level, which reveals the
synthesized Python code at the bottom of the details panel:

\begin{center}
  \includegraphics[width=0.35\textwidth]{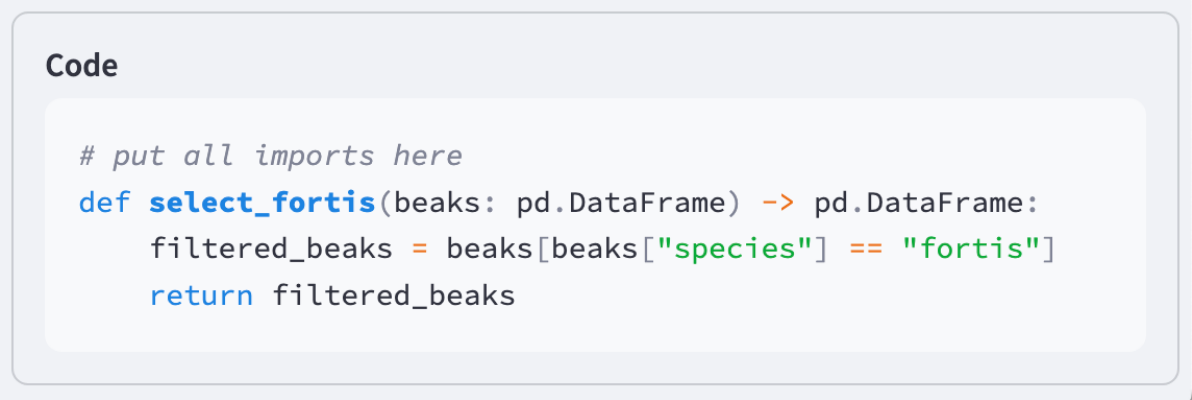}
  \Description{A screenshot of part of the details panel showing the source code 
  for the Select-Fortis node.}
\end{center}

As shown in Figure~\ref{fig:checks}, \analyst continues to \step{K} create nodes
to estimate the mean beak length for the Fortis finches using bootstrap
resampling and the percentile method.  In particular, \analyst adds
\node{Bootstrap-Average} to create an array of resampled means and then adds
\node{Estimate-Length} to compute the 95\% confidence interval for the mean beak
length.  \analyst also adds the node \node{Plot-Statistics} to plot the resampled
means for visual inspection.  After adding those nodes, \analyst runs the graph to
compute a 95\% confidence interval of approximately $10.49 - 10.64$.

Upon seeing the results, \analyst decides to make several small edits to the
dataflow graph to improve the presentation.  
Via AMA chat, \analyst adds an additional requirement to
\node{Plot-Statistics} that the original mean should be shown on the plot and
saves the change\footnote{Such changes can also be made directly rather than
through chat, as illustrated in \autoref{sec:userEdits}.}.  Outputs from
all predecessors of a node are available, so \node{Plot-Statistics} is able to
compute the original mean directly from the \node{beaks} output.  \analyst also asks
that the result of \node{Estimate-Length} be rounded to two decimal places.  The
final graph is shown in Figure~\ref{fig:checks}.

\paragraph*{Checks and Unit Tests.}
To gain confidence in the computation performed by \flowco and the results,
\analyst then adds assertions to check the node outputs in the graph.
Specifically, \step{L} \analyst switches from the Edit view to the Checks view at
the top of the Project panel and then \step{M} clicks the edit pencil icon that
appears while hovering over \node{Bootstrap-Average}. The dialog box shown in
Figure~\ref{fig:checks} appears. \analyst \step{N} adds one check manually:
``Verify that the length of the \texttt{bootstrap\_average} list is at least
5,000", and then \step{O} clicks the Suggest button.  \flowco suggests several
additional checks for that node.  \analyst saves those checks and then repeats the
process for the other nodes in the graph.

\step{P} \analyst then clicks the Check button.  \flowco synthesizes code to run
each check and verifies that they pass.  
In this case, \step{Q} \flowco reports a failure for the
\node{Bootstrap-Average} node.  indicating that only 1,000 resamples were used
by the bootstrap process.

That number of resamples was not specified by \analyst in the requirements, and
the choices made during code generation were inconsistent with the expectations
\analyst made explicit in the checks.  Upon seeing that check failure, \analyst clicks
the Fix button, leading \flowco to adjust the node's requirements and code to
address the issue.  \flowco then re-runs that node and all downstream nodes, and
all checks subsequently pass.

\analyst then selects the Test view from the Project panel and adds a
collection of unit tests to each node to gain greater confidence that the nodes
are robust to changes in their inputs, due to either changes in the data or
preceding nodes.  \analyst chooses to add some by hand and asks \flowco to suggest others.
For example, \flowco suggests the following tests for \node{Select-Fortis}:
\begin{itemize}[leftmargin=2.5em]
  \item {Test with a DataFrame containing multiple species, including `fortis'.}
  \item {Test with a DataFrame containing no `fortis' species.}
  \item {Test with an empty DataFrame.}
  \item {Test with a DataFrame where all rows are `fortis'.}
  \item {Test with a DataFrame containing NaN values in the species column.}
\end{itemize}
\analyst runs a testing pass, during which \flowco generates code to produce
appropriate inputs for each test, calls the node's code on those inputs, and
verifies the outputs are as expected.  

For \node{Bootstrap-Average} and \node{Select-Fortis}, all checks and
tests are quantitative in nature.  \flowco also supports qualitative
checks for nodes that produce visual outputs, such as a check that
``the plot uses a color overlay to distinguish between the two
species'' for \node{Histogram-Lengths}.  \flowco utilizes the LLM to
visually inspect the output plot and determined whether the check
passes or fails.

\section{Programming Model and System Architecture}

\flowco enables data scientists to effectively formulate, implement, and reason
about data science computations while leveraging LLMs throughout the process.  
Reliability and correctness are paramount in this context and necessitate the
adoption of a programming model and system whose design places sufficient
guardrails on the LLM to achieve high confidence in the correctness of the
results.
This section first presents the \flowco programming model, and then how it
utilizes LLMs while ensuring reliability.

\subsection{Key Dataflow Graph Properties}
\label{sec:model}

\flowco's \dataflow-based programming model enjoys several key properties that
make it a natural fit both for authoring data science workflows and for
interacting with LLMs. 

\smallskip\noindent\textbf{Modularity and Composability.}
Users encapsulate each analysis step as a distinct node in the \dataflow graph,
and users add explicit edges to communicate information between steps.  This
approach allows complex computations to be structured as manageable, readily
understood, and composable components.

\smallskip\noindent\textbf{Abstraction and Refinement.}
Users can specify and reason about the behavior of a node at different
levels of abstraction, from brief summaries to detailed requirements
to actual code.  
\begin{itemize}
\item
  The summary label, shown in the canvas, gives a high-level
  description to aid in organizing and understanding the \dataflow
  graph.
\item
  The requirements refine that summary to precisely capture the node's
  behavior.
\item
  The code refines those requirements to an executable form.
\end{itemize}
The user may choose which level of abstraction to work at for any
particular task and may even choose to fully ignore the lower layers
of abstraction.

\smallskip\noindent\textbf{Explicit Dependencies.} 
Nodes within a \dataflow graph are stateless and free of externally visible side
effects, meaning that each node's output depends solely on its input and that
the edges explicitly capture all dependencies between nodes. Thus, evaluation
order is derived directly from the graph structure: a node executes only after
its input nodes complete.  Moreover, a node need only recompute its output when
its inputs or implementation changes.

\subsection{System Architecture}
\label{sec:architecture}

\flowco is deployed as a web service.  It is written in Python and uses
the Streamlit~\cite{streamlit} framework to provide a web interface.
\flowco can use a variety of LLMs, but the default is OpenAI's
GPT-4o model~\cite{gpt-4o}, which has proven effective across the variety of
tasks that \flowco performs.  Further, that model's API supports
multi-modal inputs, tool calls, streaming, and structured output, all
of which \flowco uses.  \flowco's primary internal data structure is
the \dataflow graph.  The user interface is built around views of and
modifications to this graph, as are \flowco's collection of LLM agents
and mechanisms for ensuring reliability.  The most salient aspects of
those features and agents are described in the next section.

\section{Ensuring Reliability}
\label{sec:reliability}

This section outlines \flowco's mechanisms and guardrails for
ensuring reliability.

\subsection{Code Synthesis and Evaluation}
\label{sec:synthesis}

\newcommand{\pstep}[1]{\item{\bf #1.}}

\flowco adopts modular code generation to constrain the scope of any
individual LLM operation to small,
well-defined, and isolated synthesis tasks.  In particular, each LLM
synthesis step is restricted to a single node in the \dataflow graph, and
to a single abstraction layer within that node.  This modular approach
ensures that the LLM's output is limited in scope, focused and more 
easily understood than larger, more expansive code generation tasks.

\flowco's code synthesis and evaluation process for a node is divided
into three distinct steps:

\begin{itemize}[leftmargin=20pt]
\pstep{Requirements Step} The inputs to this step are the
\dataflow graph both in text-based JSON and as an
image\footnote{Evidence suggests that leveraging redundancy between
  vision and text can enhance model
  performance~\cite{DBLP:conf/nips/WangMSVW0J24}.}, as well as the
requirements for all predecessor nodes, which serve as the
preconditions for incoming values.  Nodes are processed topologically
to ensure the availability of the predecessor requirements but can
otherwise be processed in parallel.

\flowco first uses the LLM to check the consistency of the preconditions.
\flowco then generates a precise description of the node's behavior, in the form
of a prose bullet list, and also the node's output type, in the form of an
extended Python type.  These extended types capture additional information beyond what is
present in regular types. 
For example, they include the element type and meaning for a list, and
the column names, types, and descriptions for a dataframe.

\pstep{Code Step} The inputs to this step are a node's preconditions,
requirements, and output type.  \flowco uses the LLM to refine the requirements
into an executable Python function whose parameter types match the incoming data
and whose return type matches the output.  As illustrated in
\autoref{fig:codeprompt}, \flowco provides a template for this function, as well
as a pydoc documentation string including descriptions of the parameters, return
value, preconditions, and requirements.
Code generation for a node may proceed as soon as that node's requirements are
available.

\begin{figure}[t!]
  \begin{center}
  \footnotesize
  \begin{minipage}{0.475\textwidth}
  \begin{minted}{python}
# put all imports here
def select_fortis(beaks: pd.DataFrame) -> pd.DataFrame:
  """
  select fortis

  This function has the following behavior:
  - The the result output is a Pandas DataFrame containing a subset
    of rows from the `beaks` DataFrame where the species is 'fortis'.
  - The the result DataFrame retains the same columns and column 
    names as the `beaks` DataFrame.
  
  Args:
    beaks (DataFrame[
     'species': str  # The species name as a string.
     'Beak length, mm': float  # The length of the beak in millimeters.
     'Beak depth, mm': float  # The depth of the beak in millimeters.
    ]): The structure of the dataset representing bird species and 
        their beak dimensions.
      
      Preconditions:
      - beaks is the dataframe for the `beaks` dataset.
  
  Returns:
    DataFrame[
     'species': str  # The species of the bird.
     'Beak length, mm': float  # The length of the beak in millimeters.
     'Beak depth, mm': float  # The depth of the beak in millimeters.
    ]: The output is a filtered DataFrame containing only rows where 
       the species is 'fortis', retaining all original columns and 
       column names.
  """
  # put code here
  ...
  \end{minted}
  \end{minipage}
  \end{center}
  \caption{\label{fig:codeprompt}When \flowco asks the LLM to generate
    code for \node{Select-Fortis} during the compile step outlined in
    \autoref{sec:synthesis}, it provides a template containing the
    function signature and a pydoc string describing the parameters,
    preconditions, and requirements.}
    \Description{
    The code generation template for the `select_fortis' function. It
    includes a docstring with a description of the function's behavior,
    arguments, preconditions, and return value. The docstring specifies
    that the function takes a DataFrame as input, which contains columns
    for species, beak length, and beak depth. The function returns a
    DataFrame with the same structure, but filtered to include only rows
    where the species is 'fortis'. The code generation template also
    includes a comment indicating where the actual code should be
    inserted.
    }
  \end{figure}

\pstep{Evaluation Step}  The inputs to this node are 
the node's code and the input values.  
As before, evaluation proceeds in topological order and exploits
parallel evaluation whenever possible.  \flowco employs a shared pool
of Python kernels for this step.
\end{itemize}
To build and run a graph, \flowco schedules agents to perform each
individual synthesis step for each node, respecting the dependencies
between nodes and exploiting parallelism wherever possible.

\subsection{Error Detection and Repair.}
\label{sec:repair}

It is, of course, possible that the LLM fails to produce correct code.  To
mitigate synthesis failures, \flowco employs three low-level repair mechanisms during evaluation.
First, \flowco checks that a node's code is syntactically well formed and asks the LLM to
correct any syntax errors.  It provides the LLM with the details of the error
and the node's requirements, output type, and code to help the LLM understand
the context of the error.
Second, \flowco catches all run-time errors and again asks the LLM to correct
them.  It provides the LLM with details about both the node and the error in
this case.
Finally, \flowco validates the output value against the synthesized extended
output type to ensure that the output meets the node's requirements, again asking the LLM to repair the node if the
output does not match the expected type.

In all cases, \flowco first attempts a local repair to the
offending node.  If the error persists after three repair attempts,
\flowco gives the user the option to attempt a global repair via an
LLM agent that has access to the entire \dataflow graph.  
The LLM is directed to provide a minimal set of changes that fixes the problem.

\subsection{Observability and Understanding}

\flowco's \dataflow graphs, abstraction layers, and modular synthesis strategy
provide a clear and structured way to observe and understand each individual
step of the synthesis process in isolation.  In particular, each individual LLM
operation is restricted to modifying a specific part of a node, eliminating the
possibility that any other parts of that node or other nodes are inadvertently
changed.

The \dataflow graph also provides sufficient structure to enable the LLM to
readily aid the user in understanding the computation and outputs.
This is best demonstrated by the AMA chat agent, which serves as the
primary direct interface between the user and the LLM.  The AMA chat
agent equips the LLM with several capabilities to enable it to handle
a wide variety of user queries and requests. First, the LLM can
inspect the entire \dataflow graph and the contents of each node,
including output values and plots. Second, the LLM can run code
snippets in the context of the graph, allowing it to evaluate
expressions, run tests, and perform analyses that reference any part
of the graph.  Finally, the LLM can even change the \dataflow graph
directly by adding, removing, or modifying nodes and edges.  The agent
employs the tool call mechanism provided by the
LLM~\cite{openai-function-calling} to implement these capabilities,
and requests that the LLM provide an explanation of its reasoning and
rationale when performing any of these actions.

\subsection{Robust Updates} \label{sec:userEdits}

Users may iteratively refine or change the \dataflow graph either via direct edits in the canvas or via AMA chat.  Such changes may
(i) modify the graph structure by adding or removing nodes or
the edges between them; or
(ii) modify a node at any level of abstraction: description,
requirements, or code.
The former necessitates examining and possibly modifying downstream nodes to
accommodate the change.  The latter additionally necessitates examining and ensuring
consistency among the abstraction layers within the modified node.
\flowco handles these cases through a variety of techniques for change propagation
both within and between nodes.

\begin{figure}[t!]
  \centering
  \includegraphics[width=0.475\textwidth]{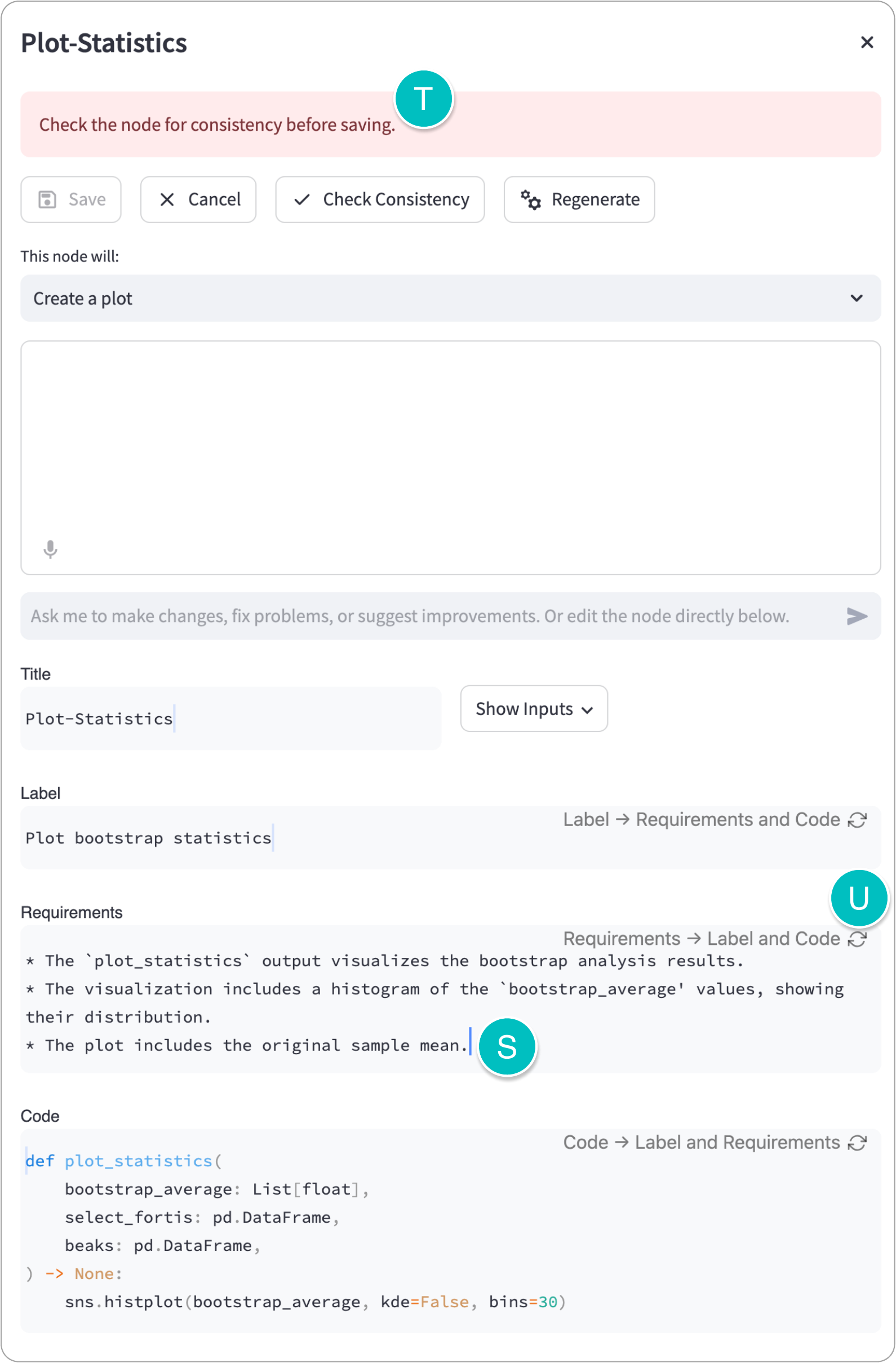}
  \caption{\label{fig:edit}The user directly modifies the components
    of a node via an editor dialog box.  The dialog box allows the
    user to directly edit the node's title, summary label,
    requirements, and code.  The user may also propagate changes in
    one component to others.  For example, \step{S} \analyst adds a
    requirement, which  \step{T} brings up a warning that the node's different components
    may not be consistent.  \step{U} The user then clicks the propagation button to
    update the summary label and code.  The editor also supports
    making modifications via chat, checking consistency between the
    components, and regenerating the components from scratch.}
  \Description{
    A screenshot of the Flowco node editor showing a dialog box for editing
    the node's components. The dialog box includes a popup for changing the
    node's type, as well as fields for editing the node's title, summary
    label, requirements, and code. Each edit field contains a button to
    propagate changes to other components.  There is also an AMA chat box for
    making changes. Across the top are buttons to save or cancel the changes,
    to check consistency between components, and to regenerate the components
    from scratch.  UI elements corresponding to specific steps in the usage scenario are
    highlighted with colored circles labeled `S' to `U'.
  }
\end{figure}

\subsubsection*{User edits.} 
The user may directly edit a node's components by clicking on the
pencil icon that appears while hovering over a node while in the edit
view\footnote{This is similar to editing the checks for a node while
  in the checks view, as shown in \step{M} of
  Figure~\ref{fig:checks}.}.  Returning to our usage scenario
in~\autoref{sec:usage-scenario}, suppose \analyst wishes modify the
\node{Plot-Statistics} node to show the original sample's mean in the
generated histogram.  \analyst clicks on the pencil icon for that node
to bring up the dialog box shown in Figure~\ref{fig:edit}\footnote{The
  code may be hidden if the user is working at the ``Requirements''
  level of abstraction.}.  \step{S} \analyst then adds ``The plot
includes the original sample mean'' to the requirements, at which point
\step{T} a warning appears indicating that the node's summary label
and code may not be consistent with the new requirements.  \step{U}
\analyst then clicks the requirement's change propagation button,
which updates the summary label and the code to reflect the new
requirements.  Once the node is returned to a consistent state, the
Save button is enabled, and \analyst can save the changes and
return to the graph editor.

The node editor provides several other features.  The Check
Consistency button uses the LLM to check the node's title, label,
requirements, and code for consistency, reporting any warnings for the
user to address.  The Regenerate button uses the LLM to regenerate
the node's components using their current values as a guide and
choosing how to resolve any inconsistencies.  Both are useful in
practice, as demonstrated in the user study presented in
\autoref{sec:user-eval}.
A node-specific AMA agent is also available to explain the node's behavior and make changes
conversationally.  Chat is often the most expedient way to make small
changes, as the AMA agent ensures consistency among the constituent
parts as it makes modifications.

\begin{figure*}[t!]
  \centering
  \includegraphics[width=0.9\textwidth]{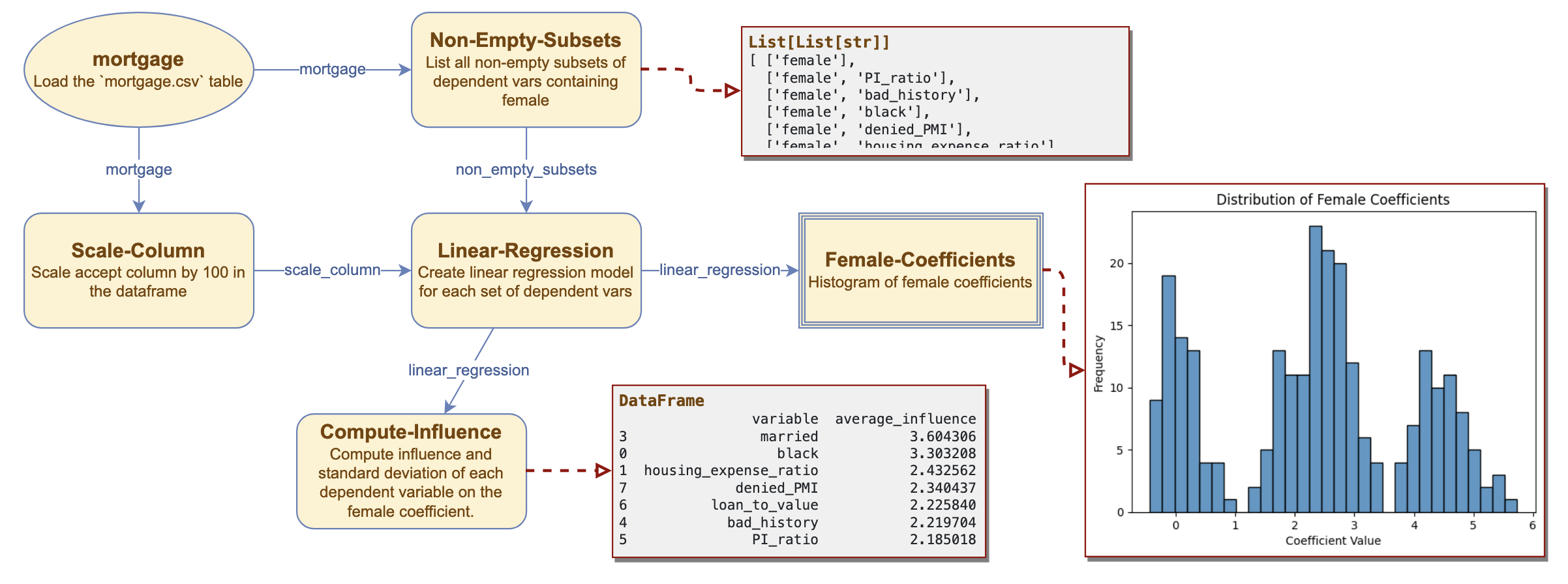}
  \caption{A multiverse analysis presented in \ref{sec:multiverse} that explores how a female applicant's
    likelihood of mortgage approval varies depending on different
    combinations of control
    variables~\cite{DBLP:journals/tvcg/LiuKAH21}.
  }
  \Description{
    A dataflow diagram showing the multiverse analysis pipeline for the
    mortgage dataset. The process begins with a node to load the data table.
    The data flows into a Scale-Column node that scales the accept column by
    100. Simultaneously, the data is passed into a Non-Empty-Subsets node,
    which generates all non-empty subsets of dependent variables that include
    "female." A table to the side shows the resulting list of subsets.

    These subsets are fed into a Linear-Regression node that creates a
    regression model for each subset. This output of those regressions flow
    into two nodes:

    1. A Compute-Influence node, which calculates the influence of each
    dependent variable on the female coefficient. A table shows results, with
    married and black showing the greatest influence.  Numerically, the
    influence measures are 3.6 and 3.3 for those variables and under 2.5 for
    the other variables.
    
    2. A Female-Coefficients node that generates a histogram of female
    coefficient values.  There are three peaks around 0, 2.5, and 4.5.
  }
  \label{fig:mortgage}
\end{figure*}

\subsubsection*{Change propagation.}
\flowco leverages the explicit dependencies captured by the \dataflow
graph structure to propagate changes through the graph during its
build process.  Specifically, when a node's incoming/outgoing edges or
requirements are modified, \flowco invalidates all downstream nodes
and resynthesizes their requirements and code under the upstream
change.  \flowco instructs the LLM to make the smallest set of
modifications necessary to address incoming changes to each node.  The
user may lock nodes they do not wish \flowco to modify while
propagating changes via the lock toggle button in the details panel
(marked as \step{R} in Figure~\ref{fig:checks}).  When a node is
locked, the LLM checks the node for consistency with upstream
modifications but does not make any changes to it.

\subsection{Validation: Assertions and Unit Tests}

To further increase confidence in the correctness of the
computation, \flowco provides several user-facing guardrails that allow
validation of node behavior and detection of errors introduced
by either the user or the LLM.

\subsubsection*{Assertion Checks.}
First, \flowco supports run-time assertions, which are simple checks on a node's
output that serve as safeguards to catch
unexpected behaviors, incorrect assumptions, or violations of key invariants.
Assertions are particularly helpful when dealing with evolving data inputs, subtle
bugs in synthesized code, or misunderstandings in analysis logic. 
The user can write assertions manually or have the LLM suggest them based on the
node's requirements and code.  

As illustrated in \autoref{sec:usage-scenario}, \flowco has a
Checks view separate from the main Edit view for
inspecting, modifying, and checking the assertions associated with each node.  This
design choice is intended to make it easy to develop and edit
\dataflow graphs without assertions getting in the way until they are desired.
Checks are expressed as a list of prose statements about the output
that may also refer to the node's inputs.
For data loading and computational nodes, the assertions are most
often quantitative statements, such as ``the output mean
is greater than 0.''  \flowco translates quantitative assertions into
Python code that performs the appropriate test.  Nodes producing
plots may also have qualitative assertions, such as ``The histogram is
bimodal.''  For those, \flowco leverages vision
capabilities of the LLM to validate whether the assertion holds.  

\subsubsection*{Unit Tests.}
Second, \flowco lets users attach unit tests to individual nodes
through a Tests view. Like assertions, these tests are expressed as
prose and may be written by the user or suggested by the LLM. Each
test defines a specific input scenario and the expected behavior of
the node under that scenario. These tests are particularly useful for
verifying that a node handles edge cases correctly, produces
meaningful results on representative inputs, and remains stable in the
face of changes upstream.  For each test, \flowco generates Python code to
create appropriate input values, run the node's code, and check the output
against the expected behavior.  As with assertions, \flowco supports both
quantitative and qualitative unit tests.

\subsubsection*{Error Reporting and Repairs.}
Together, assertions and unit tests help detect errors that arise
during analysis construction or evolve over time.  When failures occur,
\flowco reports them to the user and provides a Fix button to automatically address the problems.
\flowco uses the same repair mechanisms described earlier in
\autoref{sec:repair}.  It first attempts localized fixes within
the impacted node and then escalates to an optional global repair strategy if
necessary.

\begin{figure*}[t!]
    \centering
    \includegraphics[width=0.9\textwidth]{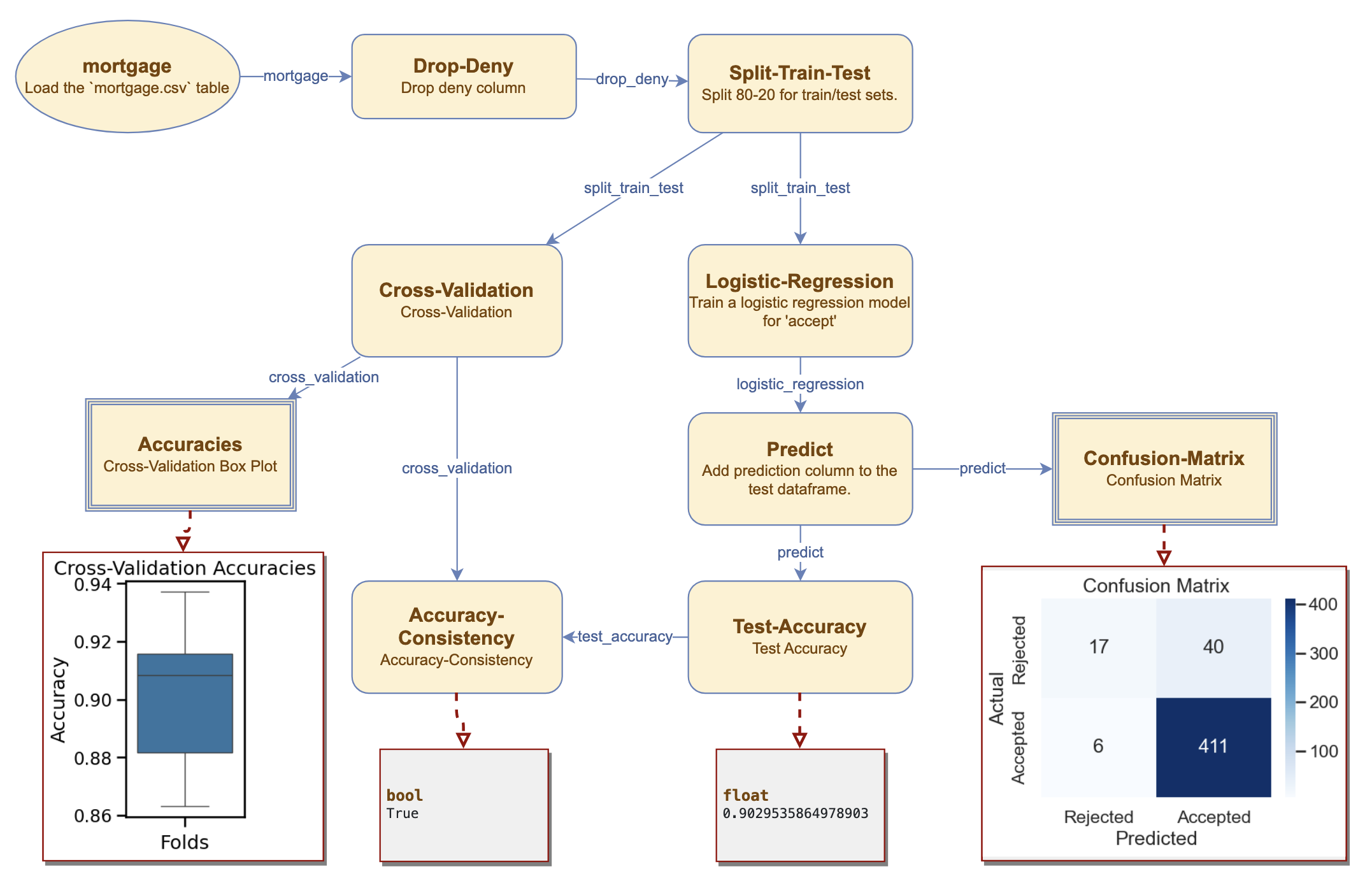}
    \caption{A logistic regression presented in \autoref{sec:logistic}
      that predicts whether a mortgage application will be accepted.
      The \dataflow graph also includes accuracy measurement and cross validation.
    }
    \Description{
    A dataflow diagram for a logistic regression model predicting
    mortgage acceptance. The flow starts with a node to loads the
    mortgage CSV file. The data is passed to Drop-Deny, which removes
    the `deny' column, and then to Split-Train-Test, which splits the
    data into 80 percent training and 20 percent testing sets.
    
    From the split, two branches emerge: 
    
    1. First branch, for model validation:    
      - The data goes to a Cross-Validation node.    
      - The results are used by:
        - Accuracies, which produces a box plot showing accuracy scores across
          folds ranging from 0.86 to 0.94, centered around 0.90.
        - Accuracy-Consistency, which confirms that accuracy is consistent
          with the second branch, outputting the value True.
    
    2. Second branch, for model training and evaluation:
      - The training data is used by a Logistic-Regression node to train a
        model on the accept label.
      - The model is used in a Predict node to add predictions to the test
        data.
      - The predicted results are evaluated using:
      - Test-Accuracy, reporting a test accuracy of approximately 0.903.
      - Confusion-Matrix, which shows: 411 correct acceptances, 6 incorrect
        rejections, 40 incorrect acceptances, and and 17 correct rejections.      
    }
    \label{fig:logistic}
\end{figure*}

\section{Example Analyses Using \flowco}
\label{sec:examples}

We demonstrate the \flowco's ability to express real-world data
analyses ranging from what may be found in introductory data science
classes to a complex multiverse analysis and a logistic regression with cross validation.

\subsection{Clustering and Linear Regression}
\label{sec:geyser}

Figure~\ref{fig:geyser} illustrates how an analyst might use \flowco to
explore the Old Faithful geyser dataset.  The dataset contains two
columns: eruptions (duration of eruption in minutes) and
waiting (waiting time between eruptions in minutes).  After loading 
the data, the analyst creates a scatter plot to visualize the
relationship between eruption duration and waiting time (\node{Duration-vs-Wait}).  The analyst
then performs k-means clustering to partition the data into two
clusters based on the eruption duration and waiting time (\node{K-Means-Clusters}), fits a
linear regression model to each cluster (\node{Linear-Regression}), and visualizes the clustered
data with regression lines (\node{Plot-Regressions}).  The analyst works entirely within the
visual editor to create the graph shown in the figure.  After viewing
the original output, the analyst makes AMA chat requests to: (1) name 
the clusters ``short'' and ``long'' and (2) adjust the colors of the final plot.

\subsection{Multiverse Analysis}
\label{sec:multiverse}

Figure~\ref{fig:mortgage} presents an analysis demonstrating \flowco's
capacity to support sophisticated analyses.  The analysis repeats a
previously published multiverse analysis on how a female applicant's
likelihood of mortgage approval varies depending on different
combinations of control variables~\cite{DBLP:journals/tvcg/LiuKAH21}.
The analyst designs a \dataflow graph to captures the key steps of that analysis,
including enumerating all combinations of control variables (\node{Non-Empty-Subsets}), fitting a
linear regression model for each combination (\node{Linear-Regression}), and plotting a
histogram capturing the variability of the estimated effect of being
female on mortgage acceptance rates and the sensitivity of that effect
to the choice of control variables (\node{Female-Coefficients}).  As in the original study, the
additional influence analysis concludes that the married and black
control variables have the greatest influence on the coefficient for
female (\node{Compute-Influence}).  \flowco enables the author to explore the multiverse of
possible models in a structured and systematic way without writing a
single line of code.

\begin{table*}[ht]
  \centering
  \begin{tabular}{llccccccc}
  \toprule
  & & \multicolumn{2}{c}{\textbf{Experience}} & \textbf{} & \multicolumn{2}{c}{\textbf{Node Creation in \flowco}} \\ 
  \cmidrule{3-4} \cmidrule{6-7}
  \textbf{PID} & \textbf{Major} & \textbf{Stats} & \textbf{Programming} & \textbf{\ \ Python Comfort\ \ } & \textbf{\ \ \ Direct\ \ \ } & \textbf{AMA} \\
  \midrule
  P1 & Data Science & 8 & 7 & Intermediate & 0 & 5 \\
  P2 & Data Theory & 7 & 7 & Intermediate & 2 & 5 \\
  P3 & Data Science & 3 & 5 & Intermediate & 1 & 7 \\
  P4 & Data Science & 6 & 6 & Intermediate & 2 & 9 \\
  P5 & Computer Science & 4 & 9 & Fluent & 5 & 0 \\
  P6 & Data Science & 7 & 4 & Not at all & 5 & 3 \\
  P7 & Data Theory & 9 & 6 & Fluent & 5 & 0 \\
  P8 & Data Science & 6 & 6 & Beginner & 5 & 1 \\
  P9 & Data Science & 9 & 8 & Very Comfortable & 7 & 0 \\
  P10 & Data Science & 5 & 4 & Intermediate & 2 & 5 \\
  P11 & Data Science/Software & 8 & 7 & Very Comfortable & 6 & 0 \\
  P12 & Data Science & 4 & 3 & Not at all & 1 & 4 \\
  \bottomrule ~\\
  \end{tabular}
  \caption{Participant background and experience. All participants are students. 
  Statistics and Programming Experiences are self-reported on a scale from 1 (very inexperienced) to 10 (very experienced).
  The ``Python Comfort'' column is the self-reported comfort with programming in Python on a scale of ``Not at all'' to ``Extremely.''
  The last two columns count the number of nodes participants created via direct editing on the canvas (Direct) or via AMA chat (AMA) during the user study (\autoref{sec:user-eval}).
  }
  \Description{
    A table showing the background and experience of the twelve participants in the study. 
    The table includes columns for participant ID, major, statistics experience, programming experience, Python comfort level, and the number of nodes created via direct editing and AMA chat.
    The participants are students in data science or related fields, with varying levels of experience in statistics and programming.
  }
  \label{tab:participant-background}
\end{table*}

\subsection{Confirmatory Analysis: Logistic Regression}
\label{sec:logistic}
The authors of the earlier multiverse
analysis~\cite{DBLP:journals/tvcg/LiuKAH21} note that logistic
regression is a better fit than linear regression for this problem
because of the binary outcome of accepting or rejecting a mortgage
application.  In Figure~\ref{fig:logistic}, the analysis applies
logistic regression to the mortgage data.  After removing the ``deny''
column (\node{Drop-Deny}), the analyst splits the data into training and testing sets (\node{Split-Train-Test}) and
performs logistic regression on all control variables 
(\node{Logistic-Regression}).  The analyst applies that model to the
testing dataset (\node{Predict}) and determines the accuracy to
be 0.903 (\node{Test-Accuracy}).  The analyst also makes a
confusion matrix for the testing dataset.

The analyst then performs a 10-fold cross validation on the training
data to construct a robust estimation of the model's expected
performance on unseen data (\node{Cross-Validation}).  The box plot
below the \node{Accuracies} node shows a median accuracy between 0.90 and 0.91,
with little variation, demonstrating the logistic model consistently
performs well across different subsets of the data.  The median is
close to the accuracy reported in the \node{Test-Accuracy} node, as expected.
The analyst adds the final \node{Accuracy-Consistency} node to make
that expectation explicit.

\section{User Evaluation}
\label{sec:user-eval}

While the examples demonstrate \flowco{}'s ability to author real-world analyses, we wanted to assess how \flowco{} could support users relatively new to data analysis.
Three research questions motivated our user evaluation of \flowco{}:
\begin{itemize}[leftmargin=20pt]
    \item \textbf{RQ1 - Authoring success.} Can participants use \flowco{} to successfully author analyses? 
    \item \textbf{RQ2 - Reactions to programming model.} What are participants' reactions to \flowco{}'s dataflow programming model?
    \item \textbf{RQ3 - User experience compared to other tools.} How does the experience of using \flowco{} compare to experiences with LLMs and other tools (e.g., scripting, computational notebooks)?
\end{itemize}

To answer these questions, we conducted a first-use study with 12 participants recruited through mailing lists, Slack groups, and professional networks related to data science. 
Participants self-reported their experience conducting statistical analyses, using LLMs for data analysis, and programming in computational notebooks.
The study participants had a wide range of analysis and domain backgrounds (see \autoref{tab:participant-background} for a summary). 
On a ten-point scale, participants ranged from 3 to 9 (mean: 6.3) for statistics experience and from 3 to 9 (mean: 6) for programming experience, with 1 as ``very inexperienced'' and 10 as ``very experienced.''
In addition, participants ranged from ``Not at all'' fluent in Python to ``Fluent,'' with the majority self-reporting as ``Intermediate,'' ``Very Comfortable,'' or ``Fluent.''
Of the twelve participants eleven are students in a Data or Statistics field and one is a student in Computer Science.

\subsection{Study Procedure}

After giving consent, the participants watched a 12-minute tutorial video on \flowco. Then,
the researcher presented participants with a dataset of airlines and their
safety~\cite{airline} and asked the participants to conduct an exploratory data
analysis for seven minutes. After the initial exploration phase, the researcher
asked participants to spend another seven minutes answering the following
question: ``To what extent can we say that fatal accidents between 1985 and 1999
predict fatality rates between 2000 and 2014?'' 
Participants were asked to think aloud while using \flowco. The seven-minute
time frames were based on pilot studies in which participants were able to use
\flowco to analyze datasets of similar complexity within seven minutes. 
When \flowco{} was not acting as expected, for example
due to bugs or LLM response latency, the researcher assisted the participants in
resolving the issue while remaining as neutral as possible. When participants
asked questions about how to use the system, the researcher provided limited
guidance to ensure all participants had as equal an experience as possible. 

Once participants arrived at a conclusion and were satisfied with their
analysis, the researcher engaged them in a semi-structured interview about their
experiences with \flowco and how the system compares to other ways in which
participants have used LLMs and computational notebooks for data analysis. 
Each study lasted approximately one hour, and participants were compensated \$25
for their time. All studies were conducted on Zoom.  Video and audio were
recorded for analysis. 

\newcommand{\participant}[1]{(#1)}

\subsection{Analysis and Findings}

We thematically analyzed researcher notes and audio transcriptions
to answer our motivating research questions.

\subsubsection{Graph creation.}
During the unguided analysis, participants created dataflow graphs
containing between five and ten nodes (median: 6.5).  Four
out of the twelve participants began by creating a summary statistics
node, such as calculating mean values across columns or asking the LLM
to generate ``summary statistics''~\participant{P1, P9, P10, P12}.
Seven of the participants created nodes via both (i) direct editing of
the graph on the canvas and (ii) AMA chat.  However, they primarily
relied on one approach or the other.  Four used direct editing
exclusively, and one used AMA exclusively.

\subsubsection{Authoring with ease.}
After the tutorial, all participants reported feeling able to use  \flowco{}
proficiently. Participants mentioned that they were able to gain this
proficiency quickly. For example, P12, who had no experience programming with
Python, expressed, ``I think it's really a good sign that I was able to learn
something like this in seven minutes or less because I'm kind of a slow learner
and it ended up being really helpful.'' Participants continued to gain confidence
as they progressed through the exploratory analysis~\participant{P1, P2, P3, P6}. Despite
some minor challenges and bugs, all participants reported that \flowco's outputs
matched their desired outcomes. 

Participants mentioned some specific tasks in \flowco that posed a
challenge to them, but only P4 and P6 encountered problems they could
not resolve.  Both were related to code generation errors that \flowco
could not repair.  These point to the need for more robust repair
mechanisms~\cite{DBLP:journals/cacm/GouesPR19,DBLP:conf/nips/YangJWLYNP24,DBLP:journals/corr/abs-2403-17134}
in the future.
Some participants, such as P8, explained, ``$\ldots$ even if I
did struggle [editing the canvas] I could have used the `Ask Me
Anything!' part to assist me.''

\subsubsection{Benefits of the dataflow graph programming model.}
Most participants found the graph-based model beneficial. Several participants 
mentioned that the graph format organized the
analysis in an easier manner to interpret compared to other tools
they have used~\participant{P3--4, P6--8, P12}.  P3 reflected, ``I think this flow is super helpful to
keep track of what you've coded and what you've done so far.''
Participants also mentioned how \flowco{} helped them understand the
process of data analysis~\participant{P4}, see how analysis steps inform each
other~\participant{P8}, and keep track of and work with multiple datasets~\participant{P7}.

Participants specifically stated that \flowco's model made it easier
to manage data and computation than computational notebooks~\participant{P3, P11}.
P3 noted that ``when you're working with pandas it just can get a
little messy $\ldots$ Between the cleaning and transforming data, and
then creating plots, creating models, it can be easy to get kind of
lost or forget where everything came from.  So I think even here being
able to see just the lines of how we got from the original dataset is
super helpful$\ldots$''

\subsubsection{Benefits of \flowco's deep LLM integration.}
Participants with experience using LLMs for data analysis said that
\flowco's integrated LLM worked better for them than using other LLM
tools~\participant{P1, P3--P7, P9, P11, P12}.  Participants mentioned
that they trusted \flowco's code more than code obtained directly from
an LLM, such as ChatGPT~\participant{P3, P5, P7}.  When asked why, P5 explained,
``Knowing that there is a layer of prompting that was carefully
engineered, and that outputs are being checked for type matching and
such, I trust this a lot more than just asking ChatGPT.''
Participants also liked not having to toggle between ChatGPT and their
tool, 
meaning that they did not have to copy and paste between the LLM and
a code editor~\participant{P1, P4, P6, P9, P11, P12}.  Other benefits of this
integration that participants noted were the speed with which they
could analyze data and the decreased effort required.  P3 described,
``I would definitely consider using this over ChatGPT $\ldots$ I'll
have to put [in] less effort.  I can just put my dataset here and then
I can visualize multiple things together.''

\subsubsection{Benefits for non-experts and data exploration.}
Participants speculated that \flowco would be 
(i) particularly beneficial for beginners and (ii) best for introductory or
exploratory data analysis~\participant{P3, P4, P6, P9, P11}.  P6 noted it would be useful in these contexts ``because it's more
visual and easier to understand.''  In addition, seven participants expressed interest in continuing to use
\flowco{} for their work~\participant{P1--P3, P5, P7, P8, P12}.  Five described a
desire to use \flowco{} over their current tools, particularly for
preliminary exploratory analysis or ``quick scratch work''~\participant{P2, P5, P3, P8, P12}.  P2
noted that they would use \flowco to ``explore ideas that I have that
I maybe I don't have time to code$\ldots$'' Additionally, participants
found that \flowco's simplicity allowed them to write their analysis
with greater ease than in other tools.  P7 described, ``I think it's
definitely less effort on the part of the person who's doing the data
analysis because you can give commands to an LLM and basically have it
do everything for you$\ldots$''

Some participants expressed concern that \flowco{} would not be able to give
correct code for more complex problems~\participant{P2, P5, P6, P9, P12}. 
This appeared to be a concern based mostly on their prior experiences using LLMs because, with the
exception of P6, these participants did not encounter issues with \flowco's code
generation.  Several participants also expressed concern that graphs with many
nodes could become ``messy'' and confusing quickly~\participant{P3, P11}.  P3 explained, ``I
would say based on the way I tend to do exploratory data analysis or just
exploring a dataset the whole flow concept here would get very messy.''

\subsubsection{LLM latency.}
Participants successfully authored analyses quickly using \flowco{} (RQ1), although
the underlying communications with LLMs introduced noticeable latency.  Seven
out of twelve participants noted this as a challenge.  For instance, P5
mentioned some confusion arising from latency, noting, ``at times I didn't know
if I was waiting on the system or if the input wasn't even picked up.'' However,
only two participants, P7 and P10, indicated that latency would likely deter
them from using \flowco{}. For others, the benefits of \flowco{} outweighed the
frustrations caused by latency. 

Advances in LLM performance~\cite{DBLP:journals/tmlr/Wan0LA0LQYZZC024} will make
latency less of an issue over time.  
In addition, \flowco can be extended to include various well-known techniques to
mitigate latency, including more broadly using asynchronous processing for LLM
requests while the user is working on other tasks, prefetching LLM responses
that are likely to be needed in the background, and employing lower-latency
models when feasible.

\subsubsection{Usability issues and fixes.}

The study revealed three specific usability issues with \flowco{}.  
First, some participants wished for an easier way to view the dataset, results,
and code~\participant{P1, P6, P7, P10}, as the right-hand panel showing that information
was not always noticeable or easily navigable.  Several \flowco features have
been refined to make that information more accessible, including more extensive
use of layovers. Second, \flowco's programming model originally restricted
users from connecting a node producing a plot to another node because plotting
nodes do not produce output values processable by downstream nodes.  This behavior was
not clear to some participants~\participant{P2, P6} and caused confusion when they wished to
structure their graphs in that way. \flowco{} now allows plot nodes to have
child nodes to match this usage pattern, and in keeping with the \flowco
dataflow model, those nodes have access to the outputs of all ancestors in the
graph. Third, several participants~\participant{P5, P9} encountered an interface refresh bug that
caused confusion over whether \flowco{} was waiting for the LLM to respond or
for the user to provide input.  That has been fixed in the current version.

\section{Discussion, Limitations, and Future Work}

The findings from the user study support the following 
answers to the research questions posed in~\autoref{sec:user-eval}.

\subsubsection*{\emph{\textbf{RQ1:}}}
Overall, participants found \flowco{} easy to learn and beneficial for
preliminary analysis.
All participants were able to successfully author analyses
independently after watching the tutorial, and the visual programming
model enabled participants to clearly organize their computation as a
graph.
Even the least experienced programmer was able to complete the
analysis presented to them.  The study also found that \flowco{} could
help more experienced participants quickly ``sketch'' analyses.  In
other words, it seems that \flowco{} not only lowers the barriers to
analyses for novices but also could equip expert users with new
capabilities.

\subsubsection*{\emph{\textbf{RQ2:}}} \flowco's deep LLM integration increased both ease of use for
participants as well as their confidence in the results.  

\subsubsection*{\emph{\textbf{RQ3:}}} After the
study, eight of the twelve participants said they would consider using
\flowco{} in the future. 
Two participants preferred \flowco{} to computational notebooks, and almost all participants preferred \flowco{} to using LLMs directly. 

\smallskip
These results suggest that \flowco{}'s mixed-initiative approach 
(i) strikes a balance between user control and LLM generation and 
(ii) provides useful abstractions and interactions for users to inspect and validate generated code. 
In this way, \flowco{} is one example of how to realize some of Horvitz's principles for mixed-initiative systems~\cite{horvitz1999principles} with LLMs.
The remainder of this section outlines a number of promising
directions for future work.

\subsubsection*{Evaluation with real-world data scientists.}

One limitation of the user study is its focus
only on data science students.  While non-experts are one of the
primary intended audiences for \flowco, that focus does limit what
conclusions can be drawn from the study.
Future work will build on these findings by evaluating \flowco across
real-world data scientists with a broader range of expertise and
skills.  Of particular interest is how \flowco performs in the context
of different domains or types of data analysis tasks, such as
hypothesis testing, modeling, or causal inference.

\subsubsection*{Statistical validity.}
\flowco{}'s architecture prioritizes the reliable generation of executable code. 
In data science, 
it is all too easy for analysts to inadvertently employ statistical
techniques or models that are inapplicable to their data or research
question.  Such mistakes can lead to invalid conclusions without
warning.  In fact, a previous study found that even expert tutorials
selected statistical tests that were inapplicable based on the
explicitly stated assumptions about data properties~\cite{jun2019tea}.
A critical line of future work is extending \flowco's guardrails to
check statistical correctness~\cite{jun2019tea,jun2022tisane}.  The
deep integration of LLMs into \flowco{} is a benefit here since LLMs
can leverage this additional guidance (e.g., as checks) to generate
statistically valid analysis code.

\subsubsection*{Hierarchical graphs.}
Participants voiced concern about how ``messy'' dataflow graphs could
become with more complex analyses.  A key challenge of visual
programming environments is managing the complexity of large
computations, especially when restricted to a single
view~\cite{DBLP:journals/csur/JohnstonHM04,suh2023sensecape}.
Techniques from hierarchical graph visualization and interaction are
an interesting avenue to explore for \flowco.  This approach may involve, for
example, supporting collapsible subgraphs, providing views that
present different parts of the graph at different levels of detail, or
changing the programming model to explicitly support hierarchical
decomposition of graphs into subgraphs that can be viewed and
manipulated independently.

\subsubsection*{Sophisticated workflow design.}
The user study demonstrates the key benefits of employing a dataflow
programming model for data science tasks.  However, \flowco presently
requires analysts to define graphs at the level of individual nodes,
which can be labor intensive for sophisticated problems.  An
interesting avenue for future work is exploring how \flowco can
support decomposition of large analysis tasks into manageable graphs
more automatically.  This may involve providing the user with
templates capturing common analysis idioms, as well as more advanced
AMA chat mechanisms that support the conversational creation of whole
dataflow graphs.

\subsubsection*{Integration with computational notebooks.}
Finally, some study participants expressed to a desire transition between
\flowco{} and computational notebooks. While \flowco{} supports
exporting analyses as Jupyter notebooks, tighter integration with
notebooks while still retaining the key benefits of \flowco remains
for future work.  Deployment studies tracking when analysts switch
modalities will inform this line of work.

\section{Conclusion}
\flowco bridges the gap between the flexibility of code and the
approachability of visual interfaces, enabling users—--especially those
with limited programming experience—--to more effectively author,
debug, and iterate on data analysis workflows. By integrating LLM
assistance throughout the development process, \flowco supports a more
interactive and exploratory approach to data science that offers a
number of clear advantages over existing authoring tools, including
computational notebooks and existing LLM-based tools. 
\flowco
additionally emphasizes correctness and transparency, providing users
with the tools to understand and validate the code generated by LLMs
and the behavior of the system.

\section{Availability}  
Source code and additional information is available at
\url{https://github.com/stephenfreund/flowco}.

\section{Summary of Author Contributions}
Stephen Freund created/contributed to the key ideas behind the work and implemented the \flowco prototype and collaborated on writing the paper.
Brooke Simon actively tested and gave feedback on the interface, contributed to the user study's design, conducted the user studies, and collaborated on data analysis and writing the paper.
Emery Berger helped develop the core ideas and guided the design of \flowco through numerous discussions and provided feedback on the paper.
Eunice Jun helped develop the core ideas and guide the design of \flowco, contributed to the user study's design and data analysis, and collaborated on writing the paper.

\begin{acks}
We thank Kyla Levin for valuable suggestions and feedback. 
We thank our participants for their time and feedback. We thank members of the UCLA Computation \& Discovery Lab for testing \flowco{}, trying out tutorials, and giving feedback on drafts of this paper. 
\end{acks}

\bibliographystyle{ACM-Reference-Format}
\bibliography{emjun, emery, steve}

\clearpage
\onecolumn
\begin{quotation}
\appendix

\section{\flowco's AMA Chat Response to ``Describe this dataset'' in \autoref{sec:usage-scenario}}\label{appendix:ama-response}

The following transcript is \flowco's response to the AMA prompt ``Describe this dataset'' in the exploratory steps of
the scenario in \autoref{sec:usage-scenario}.

\begin{quotation}

\newcommand{\h}[1]{\vspace{0.5cm}\noindent {\LARGE \textbf{#1}} \vspace{0.3cm}}
\newcommand{\hh}[1]{\vspace{0.4cm}\noindent {\Large \textbf{#1}} \vspace{0.3cm}}
\newcommand{\hhh}[1]{\vspace{0.3cm}\noindent {\large \textbf{#1}} \vspace{0.3cm}}

\sffamily
\setlength{\parindent}{0pt}

\renewcommand{\labelenumi}{\arabic{enumi}.}
\renewcommand{\labelenumii}{\alph{enumii}.}

\definecolor{mygray}{gray}{0.96}

\newenvironment{llmcode}{\begin{minted}[bgcolorpadding=0.65em,bgcolor=mygray,fontsize=\footnotesize]{text}}{\end{minted}}

\hhh{1. Restating the Question}

The question asks for a description of the dataset, which involves understanding its structure, content, and any notable characteristics. Specifically, we want to explore the columns, data types, and any patterns or distributions in the data.

\hhh{2. Approach}

To describe the dataset, I will:
\begin{enumerate}
\item Examine the structure of the dataset, including column names and data types.
\item Summarize the data to understand its size and basic statistics.
\item Check for missing values or anomalies.
\item Explore the distribution of key variables, such as \texttt{Beak length, mm} and \texttt{Beak depth, mm}.
\end{enumerate}

\hhh{3. Steps of Analysis}

I will perform the following steps:
\begin{enumerate}
\item \textbf{Inspect the structure of the dataset}: This includes column names, data types, and a preview of the data.
\item \textbf{Summarize the dataset}: Generate descriptive statistics for numerical columns.
\item \textbf{Check for missing values}: Identify if any columns have missing data.
\item \textbf{Visualize distributions}: Plot histograms or boxplots for numerical columns to understand their distributions.
\end{enumerate}

Let me start by inspecting the dataset.

\newcommand{\codeblock}[1]{\vspace{0.15cm}\noindent {\color{blue} \textbf{#1}} \vspace{0.15cm}}

\codeblock{Okay, I ran some code:}

\begin{mybox}[gray]{}
\normalsize
\begin{verbatim}
# Inspect the structure of the dataset
beaks.info()
\end{verbatim}
\end{mybox}

The dataset contains 406 rows and 3 columns. Here's the structure:

\begin{itemize}
\item \textbf{\texttt{species}}: A categorical column (object type) representing the species of the bird.
\item \textbf{\texttt{Beak length, mm}}: A numerical column (float64 type) representing the length of the beak in millimeters.
\item \textbf{\texttt{Beak depth, mm}}: A numerical column (float64 type) representing the depth of the beak in millimeters.
\end{itemize}

There are no missing values in any of the columns.

Next, I will generate descriptive statistics for the numerical columns to understand their distributions.

\codeblock{Okay, I ran some code:}
\begin{mybox}[gray]{}
\normalsize
\begin{verbatim}
# Generate descriptive statistics for numerical columns
beaks.describe()
\end{verbatim}
\end{mybox}

The descriptive statistics for the numerical columns are as follows:

\begin{itemize}
\item \textbf{\texttt{Beak length, mm}}:
    \begin{itemize}
    \item Mean: 11.35 mm
    \item Standard deviation: 1.63 mm
    \item Minimum: 8.70 mm
    \item Maximum: 15.94 mm
    \item 25th percentile: 10.20 mm
    \item Median (50th percentile): 10.90 mm
    \item 75th percentile: 11.80 mm
    \end{itemize}
\item \textbf{\texttt{Beak depth, mm}}:
    \begin{itemize}
    \item Mean: 9.12 mm
    \item Standard deviation: 0.71 mm
    \item Minimum: 7.50 mm
    \item Maximum: 10.98 mm
    \item 25th percentile: 8.60 mm
    \item Median (50th percentile): 9.10 mm
    \item 75th percentile: 9.69 mm
    \end{itemize}
\end{itemize}

These statistics suggest that the beak length has a wider range and variability compared to the beak depth.

Next, I will visualize the distributions of these numerical columns to better understand their patterns.

\codeblock{Okay, I ran some code:}
\begin{mybox}[gray]{}
\normalsize
\begin{verbatim}
# Visualize the distributions of numerical columns
import matplotlib.pyplot as plt
import seaborn as sns

# Plot histograms for beak length and depth
fig, axes = plt.subplots(1, 2, figsize=(12, 5))

sns.histplot(beaks['Beak length, mm'], bins=20, kde=True, 
             ax=axes[0], color='blue')
axes[0].set_title('Distribution of Beak Length (mm)')
axes[0].set_xlabel('Beak Length (mm)')
axes[0].set_ylabel('Frequency')

sns.histplot(beaks['Beak depth, mm'], bins=20, kde=True, 
             ax=axes[1], color='green')
axes[1].set_title('Distribution of Beak Depth (mm)')
axes[1].set_xlabel('Beak Depth (mm)')
axes[1].set_ylabel('Frequency')

plt.tight_layout()
plt.show()
\end{verbatim}
\end{mybox}

\includegraphics[width=0.45\textwidth]{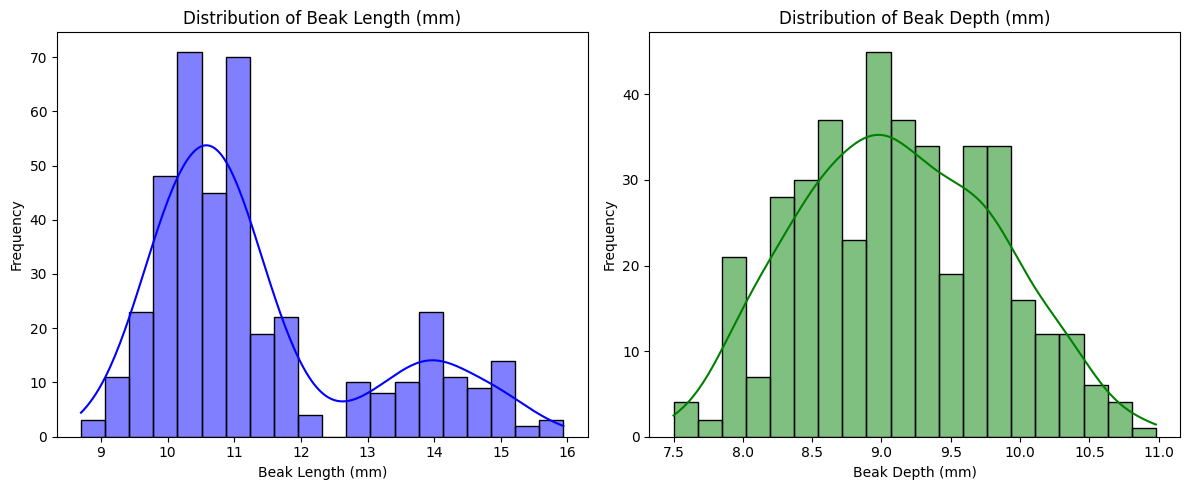}

The histograms reveal the following about the distributions of the numerical columns:

\begin{enumerate}
    \item \textbf{Beak Length (mm)}:
    \begin{itemize}
        \item The distribution appears bimodal, with two peaks around 11 mm and 14 mm. This suggests the presence of two distinct groups or species with different beak lengths.
        \item The range spans from approximately 8.7 mm to 15.9 mm.
    \end{itemize}
    \item \textbf{Beak Depth (mm)}:
    \begin{itemize}
        \item The distribution is unimodal and roughly symmetric, centered around 9 mm.
        \item The range spans from approximately 7.5 mm to 11 mm.
    \end{itemize}
\end{enumerate}

These patterns suggest that the dataset may represent multiple species with distinct beak characteristics, particularly in terms of length. The depth, however, shows less variability and a more uniform distribution.

\hhh{4. Conclusion}

The dataset contains information about bird species and their beak measurements. Key observations include:
\begin{itemize}
    \item The \texttt{species} column categorizes the data, likely corresponding to the two distinct groups observed in the beak length distribution.
    \item Beak length shows greater variability and a bimodal distribution, while beak depth is more consistent and symmetric.
\end{itemize}

This dataset is well-suited for exploring relationships between species and beak characteristics, potentially revealing ecological or evolutionary patterns.

\end{quotation}

\end{quotation}

\end{document}